\documentclass[aps,prx,twocolumn,longbibliography]{revtex4-1}
\usepackage{latexsym,epsfig,amssymb,amsfonts,amsmath,color,graphicx,bbm}
\usepackage[utf8]{inputenc}
\usepackage{hyperref}
\hypersetup{
	colorlinks   = true,
	citecolor    = black
}

\newcommand{\ii}{\textrm{i}}

\newcommand{\eqr}[1]{Eq.~(\ref{#1})}
\newcommand{\fir}[1]{Fig.~\ref{#1}}

\newcommand{\secr}[1]{Sec.~\ref{#1}}

\allowdisplaybreaks

\begin{document}

\title{Ground state phase diagram of the one-dimensional $t$--$J$ model with pair hopping terms}

\author{J. R. Coulthard$^{1}$, S. R. Clark$^{2,3}$, and D. Jaksch$^{1,4}$}
\affiliation{$^1$Clarendon Laboratory, University of Oxford, Parks Road, Oxford OX1 3PU, United Kingdom}
\affiliation{$^2$Department of Physics, University of Bath, Claverton Down, Bath BA2 7AY, United Kingdom}
\affiliation{$^3$Max Planck Institute for the Structure and Dynamics of Matter, University of Hamburg CFEL, Hamburg, Germany}
\affiliation{$^4$Centre for Quantum Technologies, National University of Singapore, 3 Science Drive 2, Singapore 117543}

\date{\today}

\begin{abstract}
The $t$--$J$ model is a standard model of strongly correlated electrons, often studied in the context of high-$T_c$ superconductivity. However, most studies of this model neglect three-site terms, which appear at the same order as the superexchange $J$. As these terms correspond to pair-hopping, they are expected to play an important role in the physics of superconductivity when doped sufficiently far from half-filling. In this paper we present a density matrix renormalisation group study of the one-dimensional $t$--$J$ model with the pair hopping terms included. We demonstrate that that these additional terms radically change the one-dimensional ground state phase diagram, extending the superconducting region at low fillings, while at larger fillings, superconductivity is completely suppressed. We explain this effect by introducing a simplified effective model of repulsive hardcore bosons. 
\end{abstract}
\maketitle

\section{Introduction}
The $t$--$J$ model has long been a subject of intense interest as a prototypical model of strongly correlated electrons because it encapsulates the physics of constrained hopping and magnetically induced real-space pairing. As such, the $t$--$J$ model has been widely studied for its relevance to high-$T_{c}$ superconductivity \cite{Zhang1988, Dagotto1994}, in particular with connection to resonating valence bond (RVB) physics \cite{Anderson1987, Lee2006}, and as a microscopic origin for the $SO(5)$ model of antiferromagnetism and superconductivity \cite{Demler2004}. Traditionally, the $t$--$J$ model emerges as an effective low-energy description of the paradigmatic Hubbard model in the limit $t \ll U$ to second order in $t/U$, where $U$ is the on-site Coulomb repulsion, giving rise to a super-exchange $J = 4t^2/U$ \cite{Chao1977, Essler2005}. The validity of the $t$--$J$ model in this context therefore necessitates $J/t \ll 1$. For high-$T_{c}$ superconductors the regime of interest is $J \sim 0.3 t$ for a two-dimensional (2D) square lattice system close to half-filling. 
 
Despite its long history, there is increasing motivation to re-examine $t$--$J$ model and explore its properties over a wider parameter space. A prominent case for this comes from the recent advances in generating strong THz fields in pump-probe experiments on solids. This technique now makes it possible to transiently manipulate materials by exciting them into non-equilibrium states not accessible thermally \cite{Kaiser2014, Singla2015, Mankowsky2016, Nicoletti2016}. Such strongly driven systems are often described by effective Hamiltonians with significant differences from those in equilibrium \cite{Mentink2014, Bukov2015, Meinert2016, Bukov2016, Mendoza2017}. In particular, the $t$--$J$ model originating from a periodically driven Hubbard model breaks the perturbative connection between $t$ and $J$, allowing $J/t$ to be controlled and the physics with $J/t > 1$ to be probed \cite{Coulthard2017}. Complementary to solid-state systems, the direct implementation of the Hubbard model and an experimental resolution of its low-temperature phase diagram is a long-standing goal of experiments with ultra-cold fermionic quantum gases in optical lattices \cite{Hart2015, Hilker2017}. In these synthetic solids, strong periodic driving, such as lattice shaking, is also routinely used to engineer the band structure \cite{Lignier2007} and microscopic interactions of the system \cite{Meinert2016}, as demonstrated recently for the super-exchange \cite{Desbuquois2017, Gorg2018}. Thus, mapping out the complete phase diagram of the $t$--$J$ model provides a fuller picture of the strongly-correlated states one might engineer by driving the Hubbard model.

\begin{figure*}
\begin{center}
\includegraphics[width=0.32\linewidth]{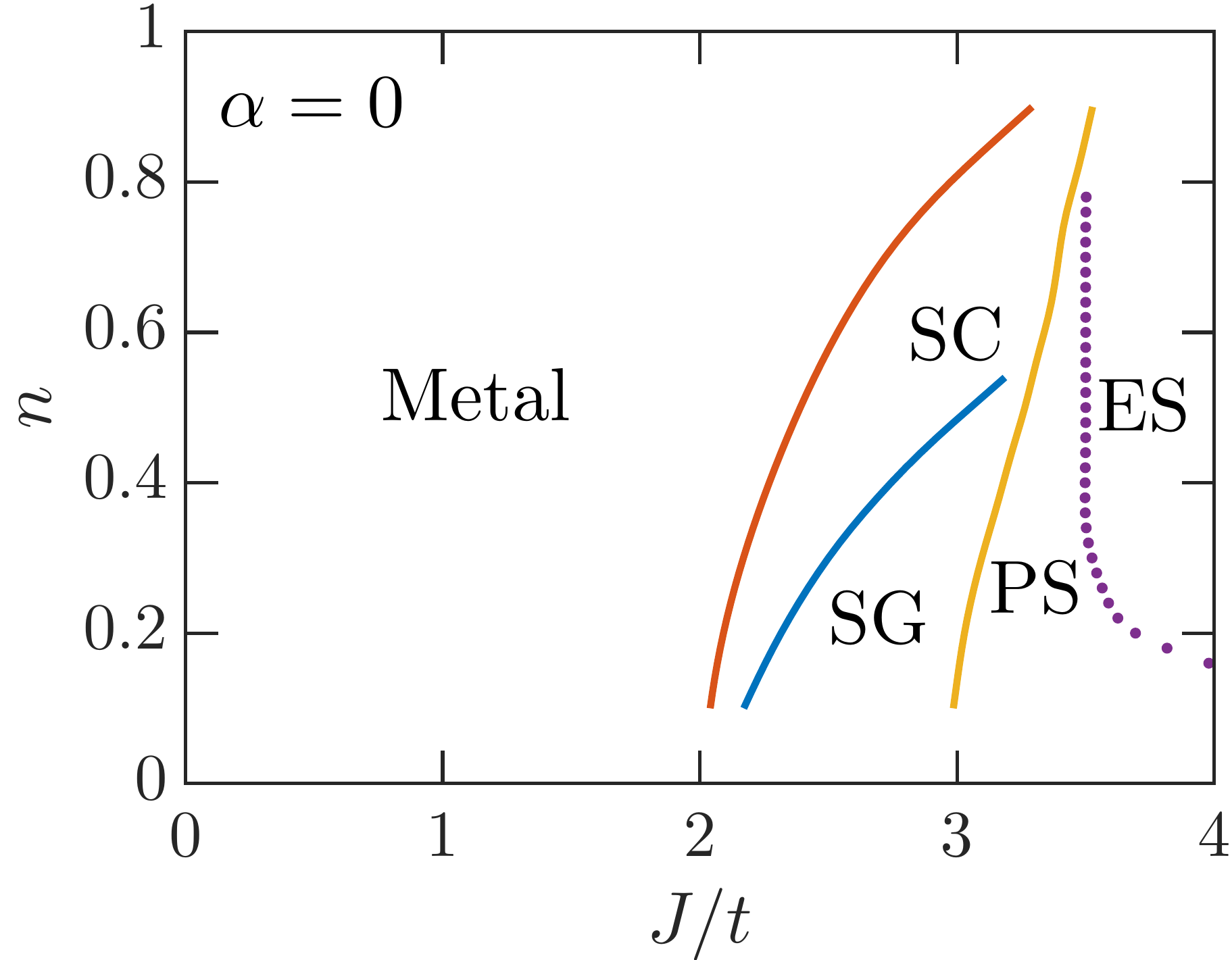}
\includegraphics[width=0.32\linewidth]{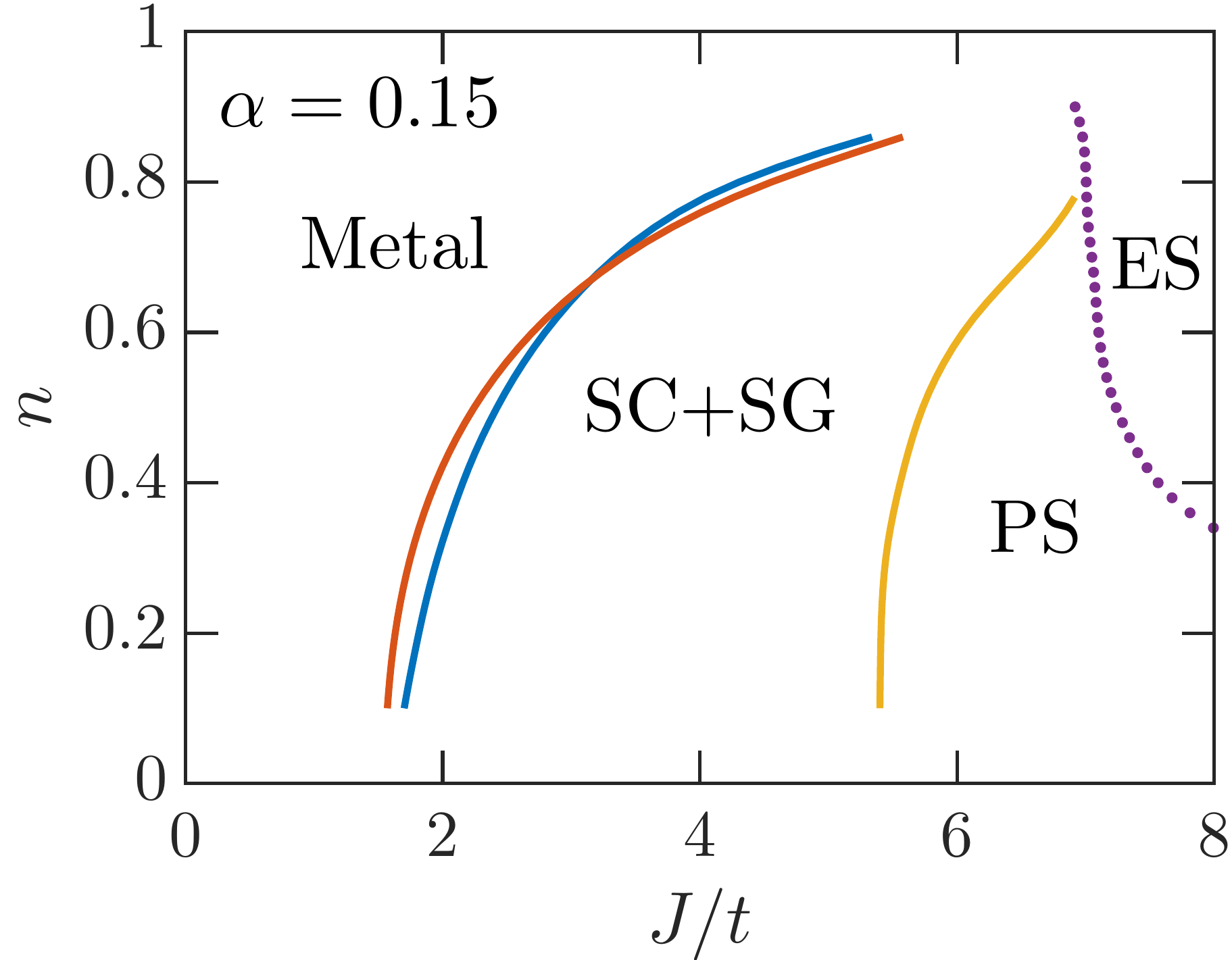}
\includegraphics[width=0.32\linewidth]{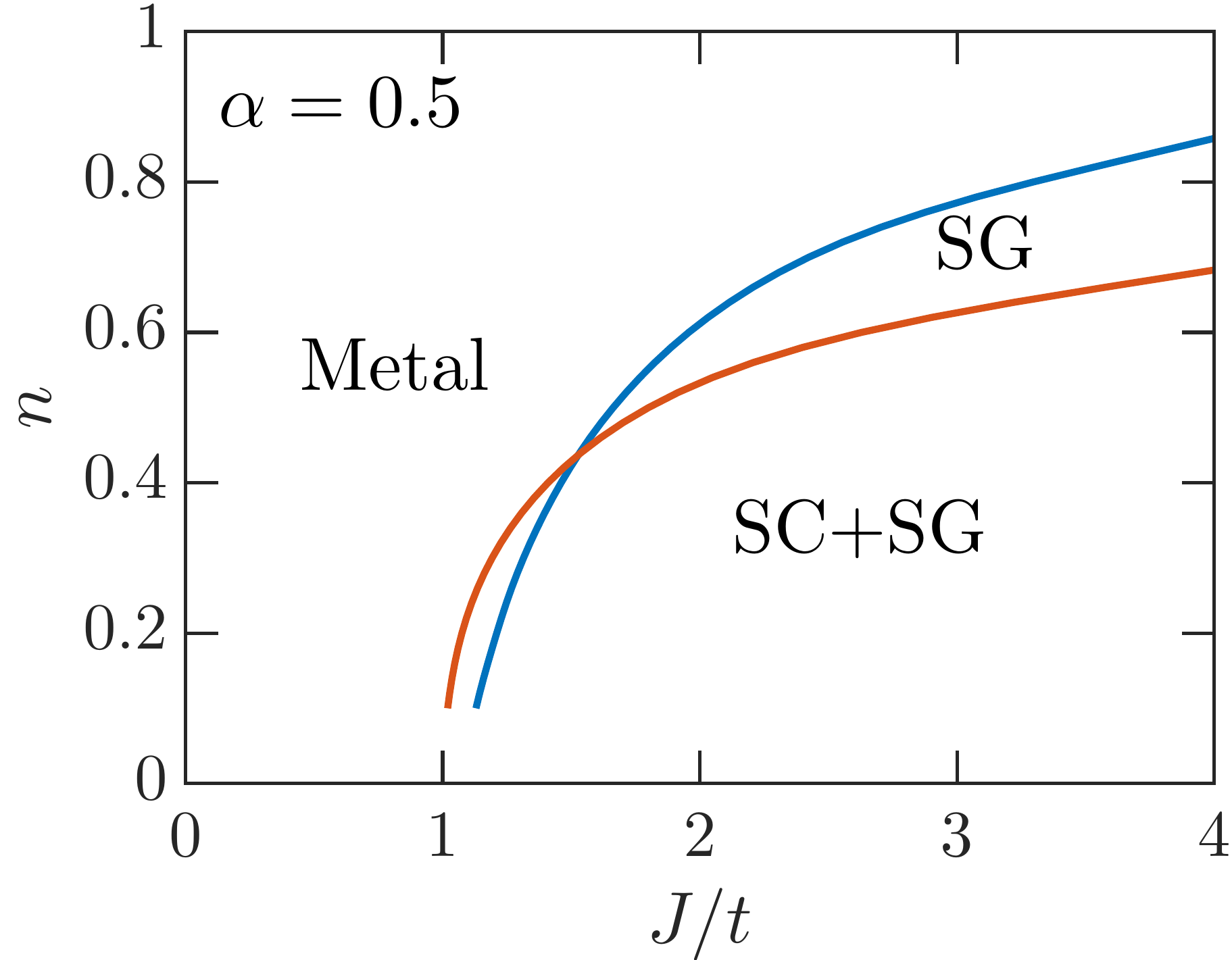}
\end{center}

\caption{Constant $\alpha$ slices through the $t$--$J$--$\alpha$ phase diagram as a function of $J/t$ and $n$. Metallic, superconducting (SC), phase-separated (PS), and electron solid (ES) phases are marked. In the $\alpha = 0.5$ case, the phase separated and electron solid phases have vanished, which we have verified for up to $J/t = 100$. \label{fig:phasediagrams}}
\end{figure*}

Motivated by these developments, in this paper we examine the ground states of the $t$--$J$ model in 1D over a wide range of $J/t$ and fillings. While much of the focus on the $t$--$J$ model is in 2D systems, the 1D system nonetheless possesses a rich phase diagram. Indeed, it displays insulating, spin-gapped and superconducting phases similar to the phenomenology of correlated materials in higher dimensions. Moreover, in 1D, density matrix renormalisation group (DMRG) \cite{White1992, Schollwock2011} provides an unprecedented ability diagnose these exotic phases in an unbiased way for large systems, allowing for accurate extrapolation to the thermodynamic limit. 

To correctly capture all the physics arising from the $t$--$J$ model with varied fillings, we crucially retain the singlet-pair-hopping term. Formally, this three-site term arises from the Hubbard model to the same order as the super-exchange $J$. Close to half-filling it is often argued that pair-hopping processes are rare \cite{Essler2005}, and so most previous studies of the $t$--$J$ model have neglected this term \cite{Ogata1991, Sano1996, Moreno2011, Corboz2014}. Some earlier works have taken pair hopping into account \cite{Spalek1987, Ammon1995, Ercolessi1997, Saiga2002, Jedrak2011}, but were restricted to mean-field approximations or used exact diagonalisations on very small systems. A key contribution of our work is that we address the $t$--$J$ model without these limitations. We find the inclusion of pair-hopping leads to a dramatically different ground state phase diagram. In particular, it has a significant impact on superconductivity by pushing the metal-superconducting boundary to lower values of $J/t$ at dilute fillings. At large fillings, the pair-hopping simultaneously increases the size of the spin-gapped region and leads to the suppression of superconductivity, in line with mean-field calculations in two-dimensions \cite{Jedrak2011}. We explain this effect by considering a simplified model of constrained hardcore bosons.

The structure of this paper is as follows. In \secr{sec:model} we introduce the $t$--$J$ model and discuss the pair-hopping term. In \secr{sec:phasediagram} we present a selection of phase-diagrams and discuss how we characterise the various phases. We then introduce in \secr{sec:effboson} a constrained hardcore boson model and compare its properties to those of the $t$--$J$ model. Finally, we conclude in \secr{sec:conclusions}.

\section{The $t$--$J$ model} \label{sec:model}
In the limit $t \ll U$ of the Hubbard model, double-occupancies are energetically suppressed. However, second-order processes, where different singly-occupied configurations are connected by virtual excitations to and from these doubly-occupied states, give rise to the $t$--$J$ model describing the effective low-energy dynamics. The $t$--$J$ model Hamiltonian may be written formally as \cite{Spalek1987} 
\begin{eqnarray} \label{eqn:tjsinglet}
\hat{H}_{tJ \alpha} &=& -t \sum_{\langle ij \rangle \sigma} ( \hat{f}^{\dagger}_{i, \sigma} \hat{f}_{j, \sigma} + \textrm{H.c.}) \nonumber \\
& &- J \sum_{\langle ij \rangle} \hat{b}^{\dagger}_{i j} \hat{b}_{i j} 
- \alpha J \sum_{\langle ijk \rangle} \left( \hat{b}^{\dagger}_{ij} \hat{b}_{jk} + {\rm H.c.} \right),
\end{eqnarray}
where $t$ is the single-particle hopping amplitude, $J = 4t^2/U$ is the strength of the super-exchange interaction, and $\alpha$ is a dimensionless constant of order unity. The definition of this model is built from projected fermionic annihilation operators for a spin-$\sigma$ fermion on lattice site $j$, defined as $\hat{f}_{j, \sigma} = \hat{c}_{j, \sigma} \hat{P}$, where $\hat{c}_{j, \sigma}$ is the corresponding canonical fermionic annihilation operator. Here $\hat{P}$ is a projector that implements the exclusion of double-occupations and is given by $\hat{P} = \prod_{j}(1-\hat{n}_{j, \uparrow}\hat{n}_{j, \downarrow})$, where $\hat{n}_{j, \sigma} = \hat{c}^{\dagger}_{\sigma, j} \hat{c}_{\sigma, j}$ is the number operator for spin-$\sigma$ fermions on site $j$. The operator 
\begin{equation*}
\hat{b}_{i,j} = \frac{1}{\sqrt{2}}(\hat{f}_{i, \downarrow} \hat{f}_{j, \uparrow} - \hat{f}_{i, \uparrow} \hat{f}_{j, \downarrow}),
\end{equation*}
annihilates a spin-singlet on lattice sites $i$ and $j$. 

The $t$--$J$ model captures two significant pieces of physics. First, it subjects the motion of electrons in a tight-binding band with hopping amplitude $t$ to a local constraint that excludes double-occupancies. Specifically, unlike $\hat{c}_{j, \sigma}$, the projection means that $\hat{f}_{j,\sigma}$ operators do \emph{not} obey the canonical fermionic anticommutation relations. This induces a non-Fermi liquid metallic state and accounts for density dependent band-narrowing effects \cite{Chao1977}. Second, neighbouring electrons experience an antiferromagnetic Heisenberg super-exchange with amplitude $J$. This induces real-space singlet pairing of electrons, which can subsequently hop with amplitude $\alpha J$, and accounts for the formation superconducting and magnetically ordered insulating states.

The parameter $\alpha$ is equal to 1/2 for a $t$--$J$ model arising from the equilibrium Hubbard model. To distinguish the Hamiltonian in \eqr{eqn:tjsinglet} from the typically studied $t$--$J$ model, which takes $\alpha = 0$, we refer to it as the $t$--$J$--$\alpha$ model from now on. Motivated by the effects of strong periodic driving on the Hubbard model we consider the regime $0 \leq J/t \leq 8$ and $0 \leq \alpha \leq 1/2$ not accessible from equilibrium. Specifically, in Appendix~\ref{sec:app_floquet} we show how periodic driving can be used to control both the single-particle hopping and the pair-hopping terms, while leaving the super-exchange unchanged.

\begin{figure*}
\begin{center}
\includegraphics[width=0.32\linewidth]{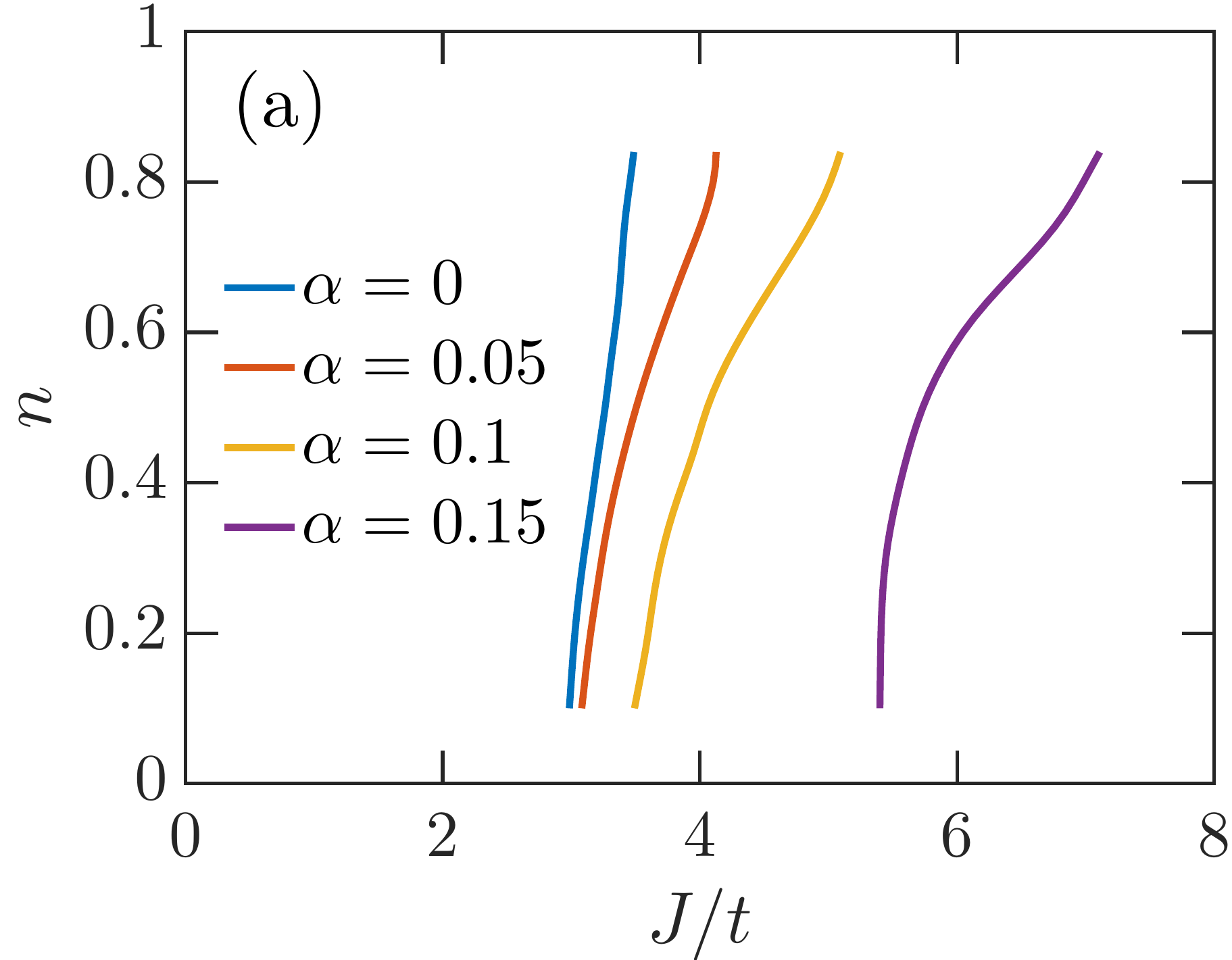}
\includegraphics[width=0.33\linewidth]{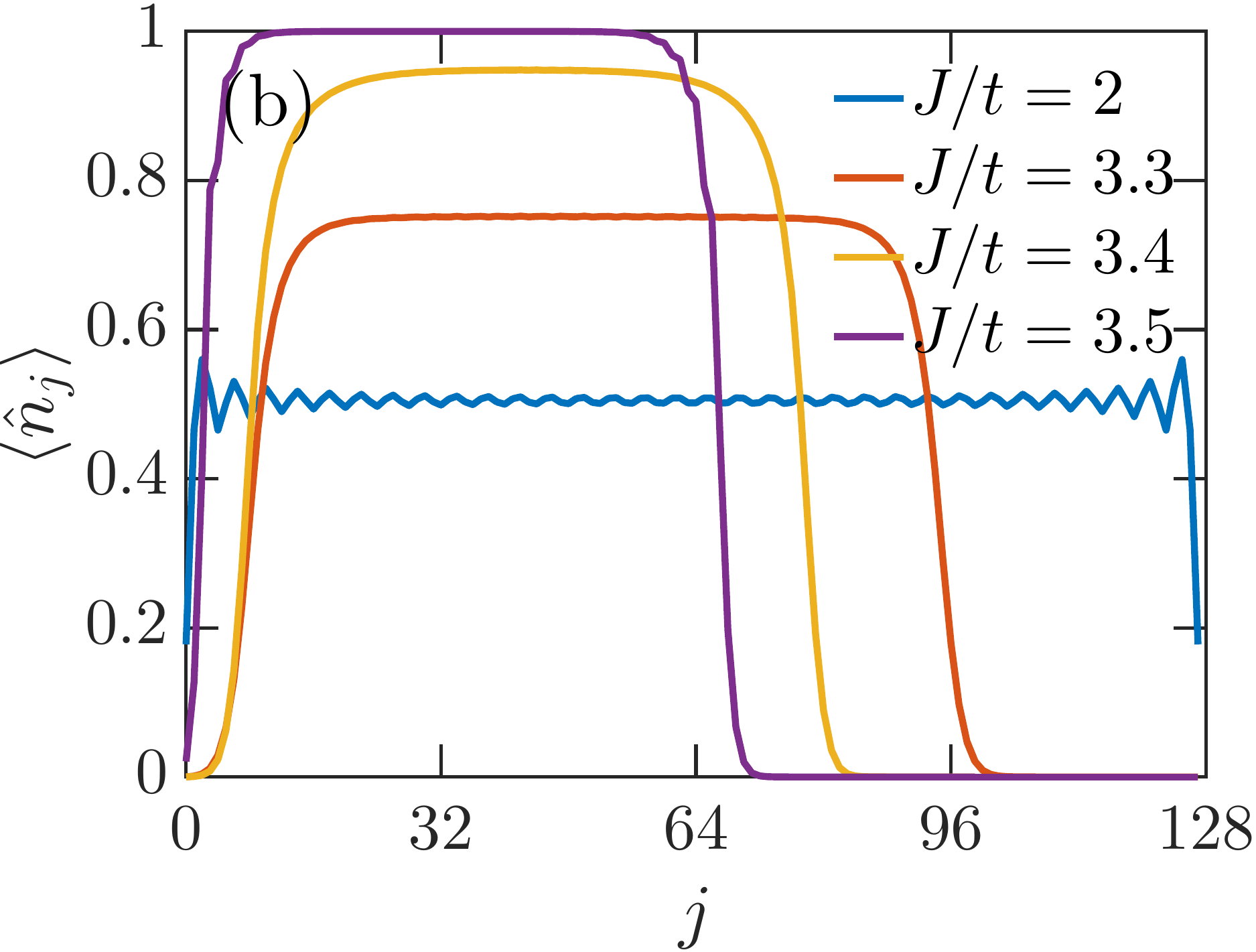}
\end{center}
\caption{(a) Phase diagram plot of $J$ vs $n$ indicating the phase separation boundary at various values of $\alpha$.
(b) Examples of the ground-state real-space fermion density $\langle \hat{n}_{i} \rangle$ for $\alpha = 0$, $n = 0.5$ on $L=128$ sites, at various values of $J/t$. \label{fig:PhaseSeparation}}
\end{figure*}
\section{Phase diagram} \label{sec:phasediagram}
To compute the ground state phase diagram of the $t$--$J$--$\alpha$ model, we use the finite system DMRG algorithm \cite{White1992, Schollwock2011} as implemented in the open source Tensor Network Theory (TNT) library \cite{Alassam2017}. Further details of the DMRG calculation are provided in Appendix~\ref{sec:dmrg_details}. We consider a 1D chain of $L$ sites containing a number of ``up" and ``down" fermions $N_{\uparrow}$ and $N_{\downarrow}$ respectively, where $N_{\sigma} = \langle \sum_{j} \hat{n}_{j, \sigma} \rangle$, and $\langle \cdot \rangle$ denotes the expectation value with respect to the ground state. Fixing this filling results in a mean number of fermions per site $n = (N_{\uparrow} + N_{\downarrow})/L$. Note that except when determining the spin-gap in \secr{sec:scspingap}, we take $N_{\uparrow} = N_{\downarrow}$. 

The main correlation functions of interest are the density-density correlations $$N_{ij} = \langle \hat{n}_i\hat{n}_j\rangle  -\langle \hat{n}_i\rangle \langle \hat{n}_j\rangle\,,$$ with $\hat{n}_{j} = \hat{n}_{j \uparrow} + \hat{n}_{j \downarrow}$, the spin-spin correlations $$S_{ij} =  \langle \hat{S}^z_i\hat{S}^z_j\rangle\,,$$ with $\hat{S}^z_i = (\hat{n}_{i,\uparrow} - \hat{n}_{i,\downarrow})/2\,,$ and the nearest-neighbour singlet-paring correlations $$P_{ij} =  \langle \hat{b}^\dagger_{i,i+1}\hat{b}_{j,j+1}\rangle\, .$$We also compute the corresponding structure factors, i.e.\ the Fourier transforms of these quantities, 
\begin{equation} \label{eq:structurefactor}
X(q) = \frac{1}{L} \sum_{jk} X_{jk} {\rm e}^{{\rm i} q(j-k)},
\end{equation}
where $X$ is any of $N$, $S$, or $P$.

Our main results, the phase diagrams for the $t$--$J$--$\alpha$ model in the $n$--$J/t$ plane, are presented in \fir{fig:phasediagrams}. For $\alpha = 0$, we reproduce the results of Moreno {\em et al} \cite{Moreno2011}. We also show the phase diagrams for $\alpha = 0.15$ and $\alpha = 1/2$ respectively, mapping the full range of phases induced by the pair-hopping. Contrary to the small exact diagonalisation results of Ammon {\em et al} \cite{Ammon1995}, we find that superconductivity does not survive at all filling fractions. We do, however, find that the spin gap extends outside of the superconducting region, indicating a gas of preformed pairs. 

In the following subsections, we explain in detail how the various phases are characterised with the above correlations and the physics at play in each of the phases.

\begin{figure*}
\includegraphics[width=0.32\linewidth]{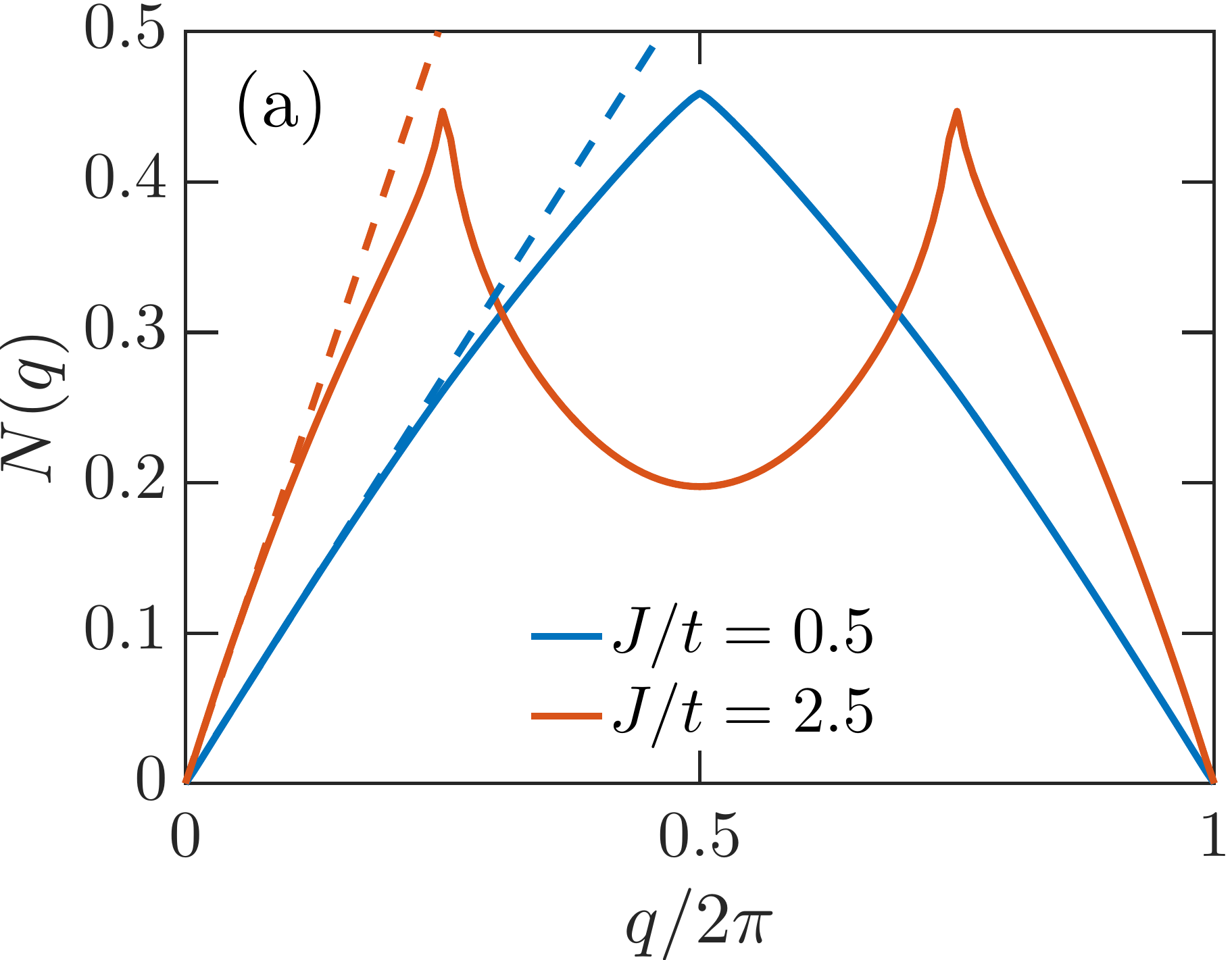}
\includegraphics[width=0.32\linewidth]{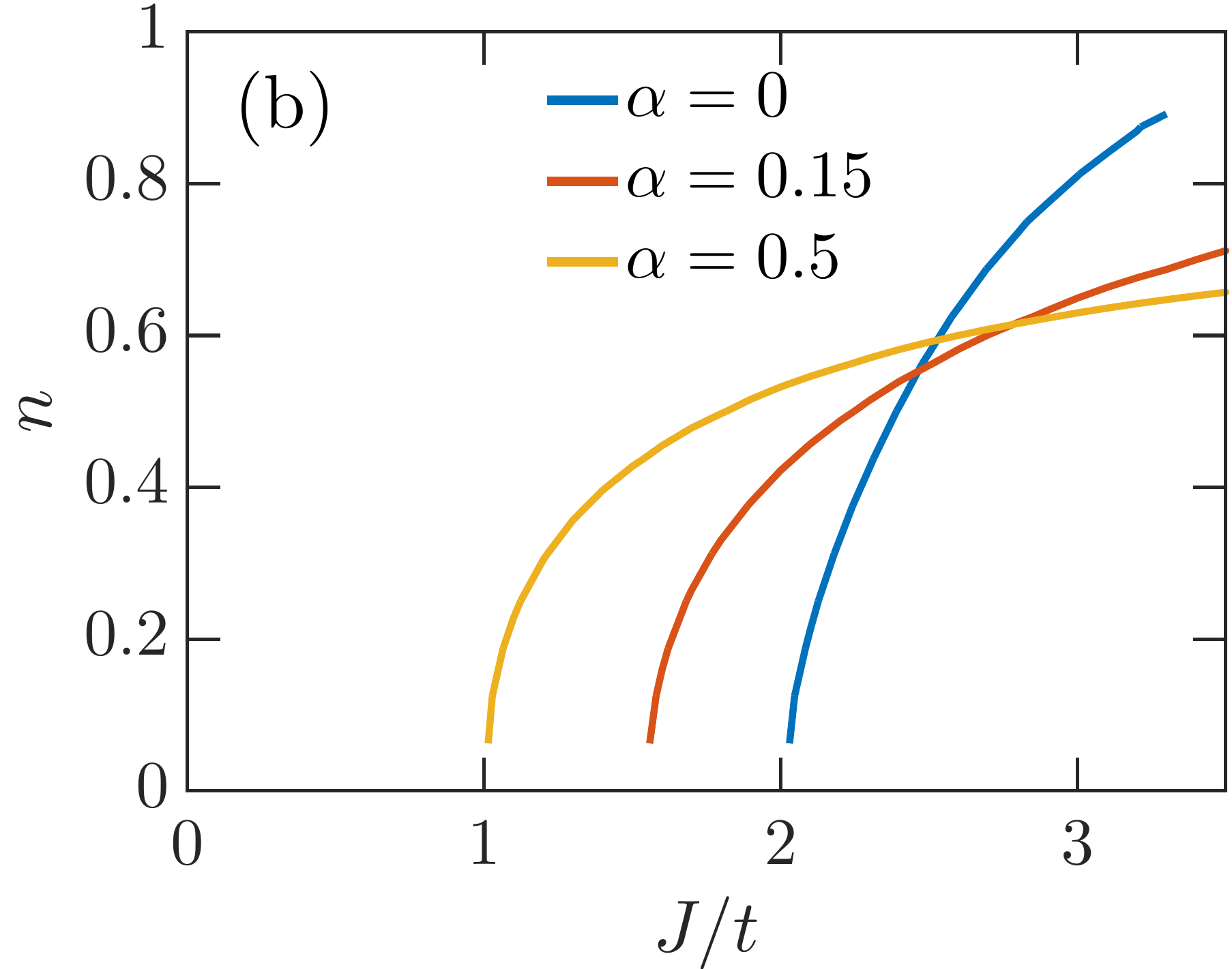}
\includegraphics[width=0.33\linewidth]{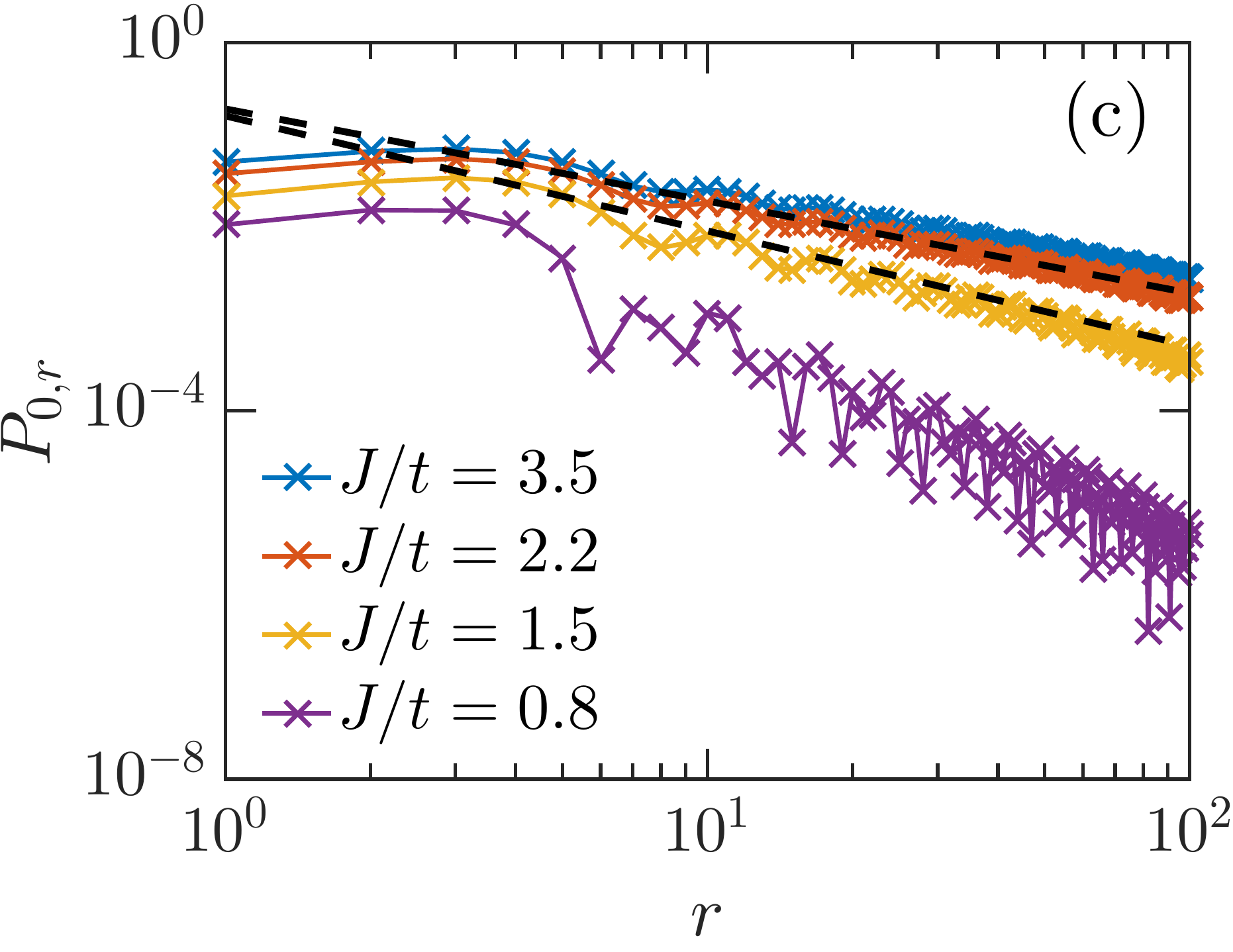}
\caption{(a) The density structure factor $N(q)$ for $\alpha = 0.5$, $L=128$ sites. The dotted lines show the linear fit as $q \rightarrow 0$, from which $K_{\rho}$ is extacted. (b) $K_{\rho} = 1$ contours for $\alpha = 0$, $0.15$, and $0.5$, extrapolated to the thermodynamic limit. (c) The real-space singlet correlations $P_{0,r}$ are marked with crosses, with a solid line as a guide for the eye. Computed for $L = 128$, $\alpha = 0.5$, $n = 0.5$, at various values of $J/t$. The upper and lower black dotted lines indicate the power law decays $0.19r^{-1}$ and  $0.16r^{-1.25}$ respectively. 
\label{fig:sc}}
\end{figure*}

\subsection{Phase separation}
The second term in \eqr{eqn:tjsinglet} is commonly rewritten as the antiferromagnetic Heisenberg coupling,
\begin{equation} \label{eqn:heisenberg}
\hat{H}_{\rm Heis} = \frac{J}{2} \sum_{\langle i j \rangle} \left( \hat{S}^{x}_{i} \hat{S}^{x}_{j} +  \hat{S}^{y}_{i} \hat{S}^{y}_{j} + \hat{S}^{z}_{i} \hat{S}^{z}_{j} - \frac{\hat{n}_{i} \hat{n}_{j}} {4}\right),
\end{equation}
where 
$\hat{S}^{x}_{j} = (\hat{f}^{\dagger}_{j, \downarrow}\hat{f}_{j, \uparrow} + \hat{f}^{\dagger}_{j, \uparrow}\hat{f}_{j, \downarrow})$,
$\hat{S}^{y}_{j} = \ii (\hat{f}^{\dagger}_{j, \downarrow}\hat{f}_{j, \uparrow} - \hat{f}^{\dagger}_{j, \uparrow}\hat{f}_{j, \downarrow})$, 
and $\hat{S}^{z}_{j} = (\hat{n}_{j, \uparrow} - \hat{n}_{j, \downarrow})$ are the spin-$1/2$ Pauli operators acting on the spin degree of freedom at site $j$. 
When written in this form, we anticipate that in the absence of pair-hopping, the $t$--$J$ model will exhibit competition between the delocalising effect of the single-particle hopping $t$ and the attractive Heisenberg-like interaction $J$. When $t \ll J$, we expect this attractive interaction to dominate, and the fermions to separate into antiferromagnetic clusters and hole-rich regions. This is known as phase separation \cite{Emery1990}. To quantitatively characterise the transition boundary, we compute the inverse compressibility,
\begin{eqnarray}
\kappa^{-1} (n) & = & n^2 \frac{\partial^2 E_{0}(n)}{\partial n^2} \nonumber \\
                & \approx & n^2 \frac{\left[ E_{0}(n + \Delta n) + E_{0}(n - \Delta n) - 2 E_{0}(n) \right]}{(\Delta n)^2},
\end{eqnarray}
where $E_{0}(n)$ is the ground state energy of the system at a filling $n$. At the onset of phase separation, the compressibility diverges, and so $\kappa^{-1}$ crosses zero. The phase separation boundary is shown in \fir{fig:PhaseSeparation}(a) for selected values of $\alpha$. We see clearly that the phase separation is suppressed with increasing $\alpha$. We further find that phase separation disappears completely for $\alpha = 1/2$ \cite{Ammon1995}. 

Where phase separation does occur, and if $J/t$ is sufficiently large, the system can become fully separated into an particle-rich region with $\langle \hat{n}_{j} \rangle \approx 1$, and a hole-rich region with $\langle \hat{n}_{j} \rangle \approx 0$. This is illustrated in \fir{fig:PhaseSeparation}(b) for $\alpha = 0$. In this plot, $J/t = 2$ is metallic, while the rest are phase-separated, and $J/t = 3.5$ indicates an electron solid phase with regions of $\langle \hat{n}_{i}\rangle = 1$, and $\langle \hat{n}_{i}\rangle = 0$. 

As noted in Ref.~\cite{Moreno2011}, we find that the phase-separated phase presents a number of issues for the DMRG calculation. Firstly, the antiferromagnetic island is off-centre for larger $J/t$. This is because the phase-separated ground state is highly degenerate, i.e. ignoring boundary effects, the cluster of fermions has very nearly the same energy regardless of where it is located in the lattice \footnote{In our results, the fermions tend to cluster on the left because the DMRG algorithm always finishes with a leftward minimisation sweep, breaking the reflection symmetry of the system.}. Relatedly, we find that for large systems DMRG encounters metastability issues deep in the phase-separated phase. The results for $J/t = 3.5$ in \fir{fig:PhaseSeparation}(b) are therefore not expected to be quantitatively representative of the true ground state (as can be seen by the lack of reflection symmetry in the antiferromagnetic cluster). Because of this, a full extrapolation of the electron solid phase boundary to the thermodynamic limit is not possible. Rather, in \fir{fig:phasediagrams}, we show the approximate boundary as the contour where $\max_{j} \langle \hat{n}_{j} \rangle > 0.999$ for $L=128$ as a dotted line.


\subsection{Superconducting region and spin gap} \label{sec:scspingap}
To identify the surperconducting phase boundary, we appeal to the Luttinger liquid formalism. When the ${t\textrm{--}J\textrm{--}\alpha}$ model is not phase-separated, it can be mapped onto either a Tomonaga-Luttinger liquid (TLL) with gapless spin and charge excitations, or Luther-Emery liquid (LEL) with a spin-gap \cite{Moreno2011, Giamarchi2004}. The central quantity in both of these models is the Luttinger parameter $K_{\rho}$. For $K_{\rho} < 1$, the TLL/LEL has repulsive interactions, whereas for $K_{\rho} > 1$, the TLL/LEL has attractive interactions, and thus superconducting correlations dominate.

We extract $K_{\rho}$ by computing the ground-state density structure factor, and exploiting the linear dependence at small $q$ values, which we show in \fir{fig:sc}(a). The linear dependence is given by \cite{Moreno2011}
\begin{equation}
N(q) \approx \frac{K_{\rho} |q|}{\pi} \quad {\rm as}, \quad q \rightarrow 0,
\end{equation}
for both the TLL and LEL. By performing a linear fit for small values of $q$, we obtain a value of $K_{\rho}$ at a given system size $L$. By computing this as a function of $L$, extrapolating to the $L \rightarrow \infty$ limit, and finding where $K_{\rho} = 1$, we determine the superconducting phase boundary, which we show in \fir{fig:sc}(b) for a few values of $\alpha$. We see clearly that the effect of the pair hopping is to shift the metal--superconducting phase boundary to lower values of $J/t$ at small fillings, and suppress SC at large fillings. We expect this observation to persist in higher dimensions, and indeed is corroborated by two dimensional renormalised mean-field theory studies \cite{Jedrak2011}. In \fir{fig:sc}(c), we show some examples of the real-space singlet correlations $P_{0,r}$ at low fermion densities. Between $J/t = 0.8$ and $J/t = 0.5$, as the system enters the spin-gapped region, we we see a clear change in behaviour as $P_{0,r}$ goes from oscillatory and rapid (but still algebraic) decay to a much slower decaying behaviour with suppressed oscillations. The changes are indicative of a gapless metal, to spin-gapped metal, to spin-gapped superconductor transition. 

The spin gap is defined as the energy gap between the “singlet" ground state, and the lowest-lying triplet excitation,
\begin{equation}
E_{\rm SG} = E_{0}(S^{z} = 1) - E_{0}(S^{z} = 0),
\end{equation}
where $S^{z} = (N_{\uparrow} - N_{\downarrow})/2$. In any finite system $E_{\rm SG}$ will be finite, vanishing only in the thermodynamic limit. It also closes rather slowly as a function of system size, and so it is again important to extrapolate to ${L \rightarrow \infty}$ \cite{Moreno2011}. The contours drawn in \fir{fig:phasediagrams} are for $E_{\rm SG}(L \rightarrow \infty) < 0.005$. 

\begin{figure*} [t!]
\begin{center}
\includegraphics[width=0.33\linewidth]{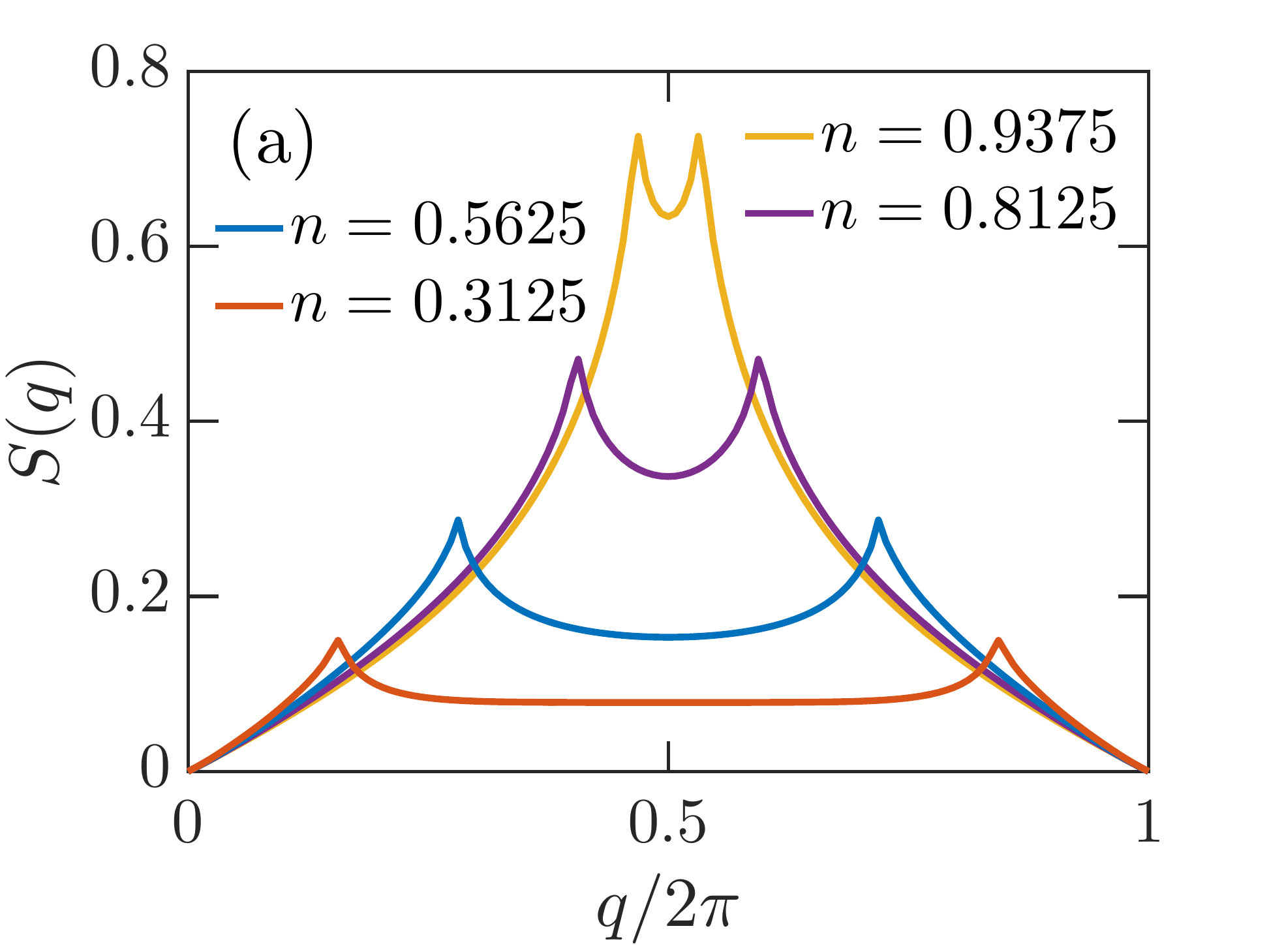}
\includegraphics[width=0.33\linewidth]{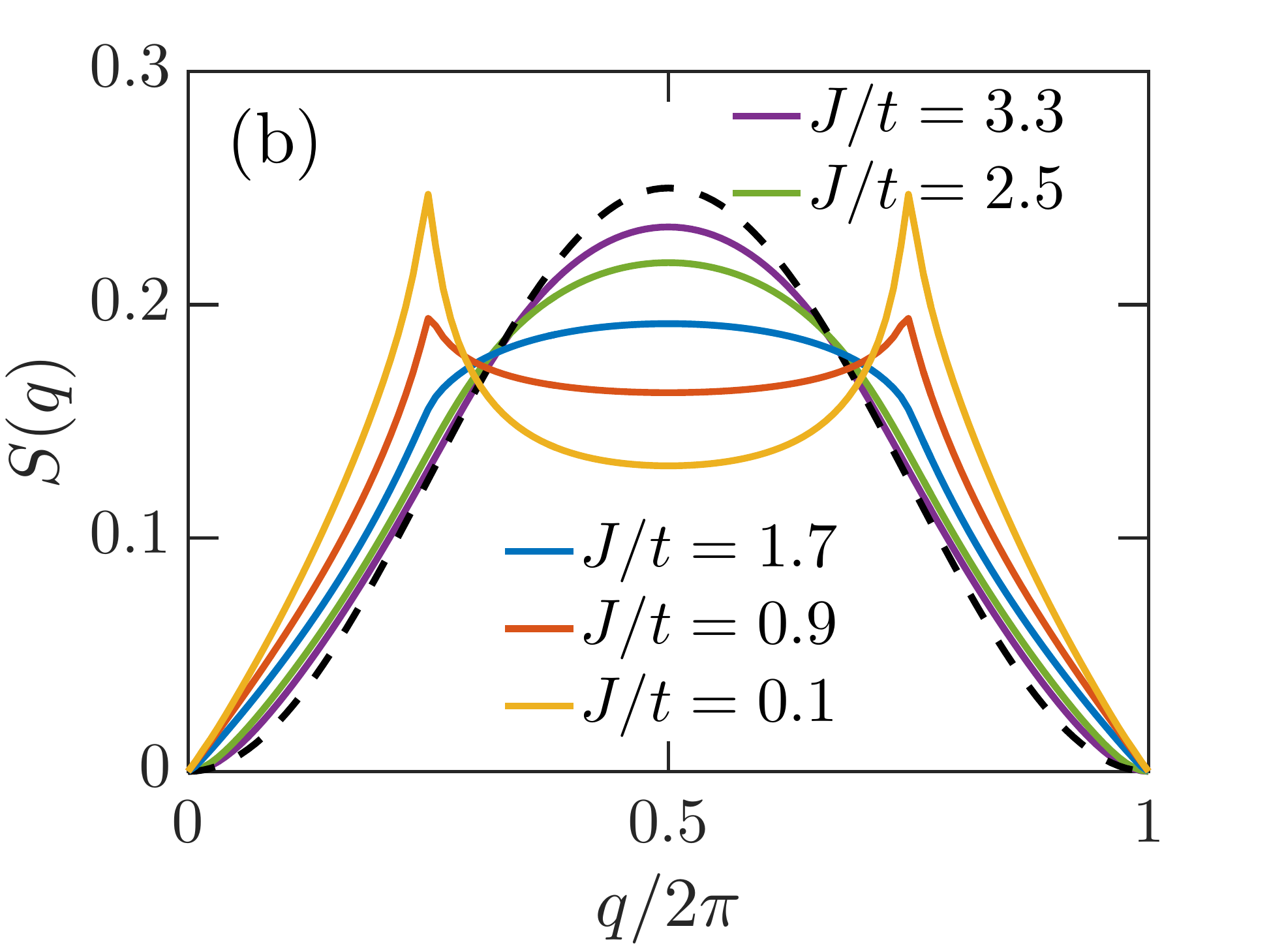}

\end{center}

\caption{Spin structure factor $S(q)$ for $L=128$ sites with $\alpha = 0.5$. In (a) $S(q)$ at $J/t = 0.1$ at various fillings $n$ is shown, while (b) shows $S(q)$ for $\alpha =1/2$ and $n = 0.5$ at various values of $J/t$. For reference, the black dashed line indicates $(1/8)(1-\cos(q))$, the structure factor for a free gas of bound pairs.
\label{fig:Magnetic}}
\end{figure*}

As we shall discuss in more detail in \secr{sec:effboson}, the presence of a finite spin-gap alters the nature of the superconducting ground state, and it becomes possible to think of the superconductor as a Bose condensate of locally bound singlet pairs. Luttinger liquid theory predicts that the long range behaviour of the singlet correlations will be
\begin{equation}
P_{0,r} \sim r^{-(1 + 1/K_{\rho})}, \qquad {\rm and} \qquad P_{0,r} \sim r^{-1/K_{\rho}},
\end{equation}
in the TLL and LEL respectively \cite{Giamarchi2004}. For reference, we indicate two algebraic decays in \fir{fig:sc}(c). The lower line given by $0.16r^{-1.25}$, while the upper line, given by $0.19r^{-1}$, indicates superconducting correlations. The real-space singlet correlations for $J/t = 2.2$ and $J/t = 3.5$ are therefore consistent with a superconducting LEL with $K_{\rho} \gtrsim 1$.

\subsection{Magnetic correlations}

From the Heisenberg term \eqr{eqn:heisenberg}, it is clear that the superexchange interaction will induce antiferromagnetic correlations in the ground state. At precisely half-filling, i.e. $n = 1$, the fermions become completely immobile and we are left only with the spin degree of freedom, which is governed by the Heisenberg Hamiltonian with an antiferromagnetic ground state \cite{Essler2005}.

Away from half-filling, the presence of holes obscures the underlying magnetic order as the single particle and pair hopping delocalise the spins. Rather than antiferromagnetic correlations (i.e. a spin wave with quasimomentum $q = \pi$), the ground state contains a spin wave with longer wavelength $q = n \pi$. This is identified by the location of the peak in the spin structure factor, which is shown in \fir{fig:Magnetic}(a) for various fillings. In \fir{fig:Magnetic}(b) we show the spin structure factor at various $J/t$. Upon entering the spin-gap, the sharp peaks, which suggest quasi-long range magnetic order, are suppressed and are instead replaced by a broad peak at $q = \pi$. This is readily understood by considering the spin-gapped phase as a gas of bound singlet pairs, which we will discuss in more detail in \secr{sec:effboson}. Each singlet pair's spin degree of freedom is maximally entangled, and so due to the monogamy of entanglement, the constituent fermions cannot have any spin correlations beyond their adjacent partner. Hence this peak has the approximate form $S(q) \sim 1-\cos(q)$, which is the form given by a free gas of antiferromagnetically bound pairs \cite{Ammon1995}, and which we indicate as a black dotted line in \fir{fig:Magnetic}(b).

\begin{figure}[t!]
\begin{center}
\includegraphics[width=0.9\linewidth]{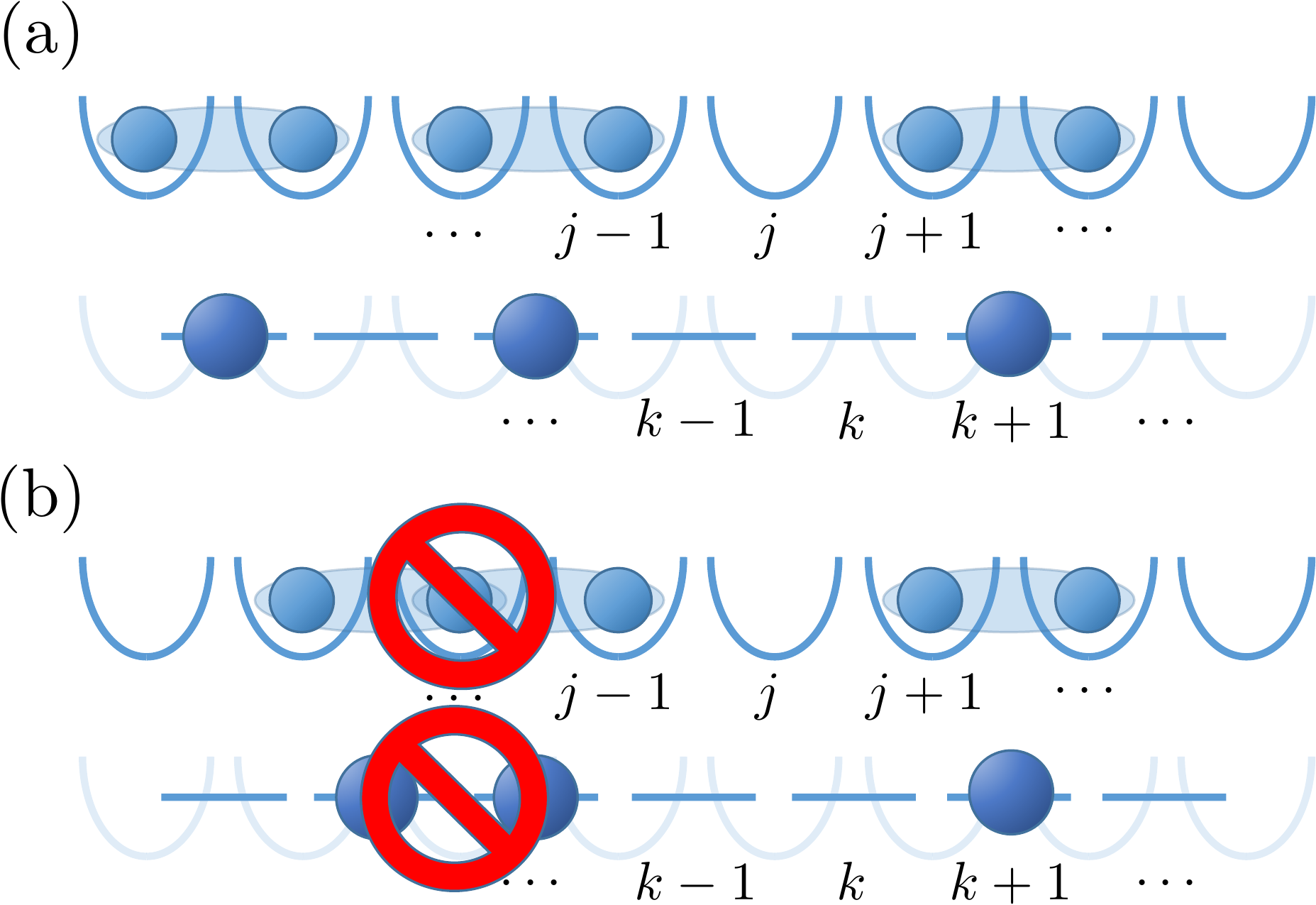}
\end{center}
\caption{(a) Schematic showing how a singlet pair of fermions may be represented in out effective model by a single hardcore boson. (b) Schematic showing how a “no-nearest-neighbour" constraint in the effective boson model arises from the “no-double-occupancy" constraint of the $t$--$J$--$\alpha$ model.
\label{fig:BosonModel}}
\end{figure}

\section{Effective Bosonic Model} \label{sec:effboson}
\begin{figure*}
\begin{center}
\includegraphics[width=0.32\linewidth]{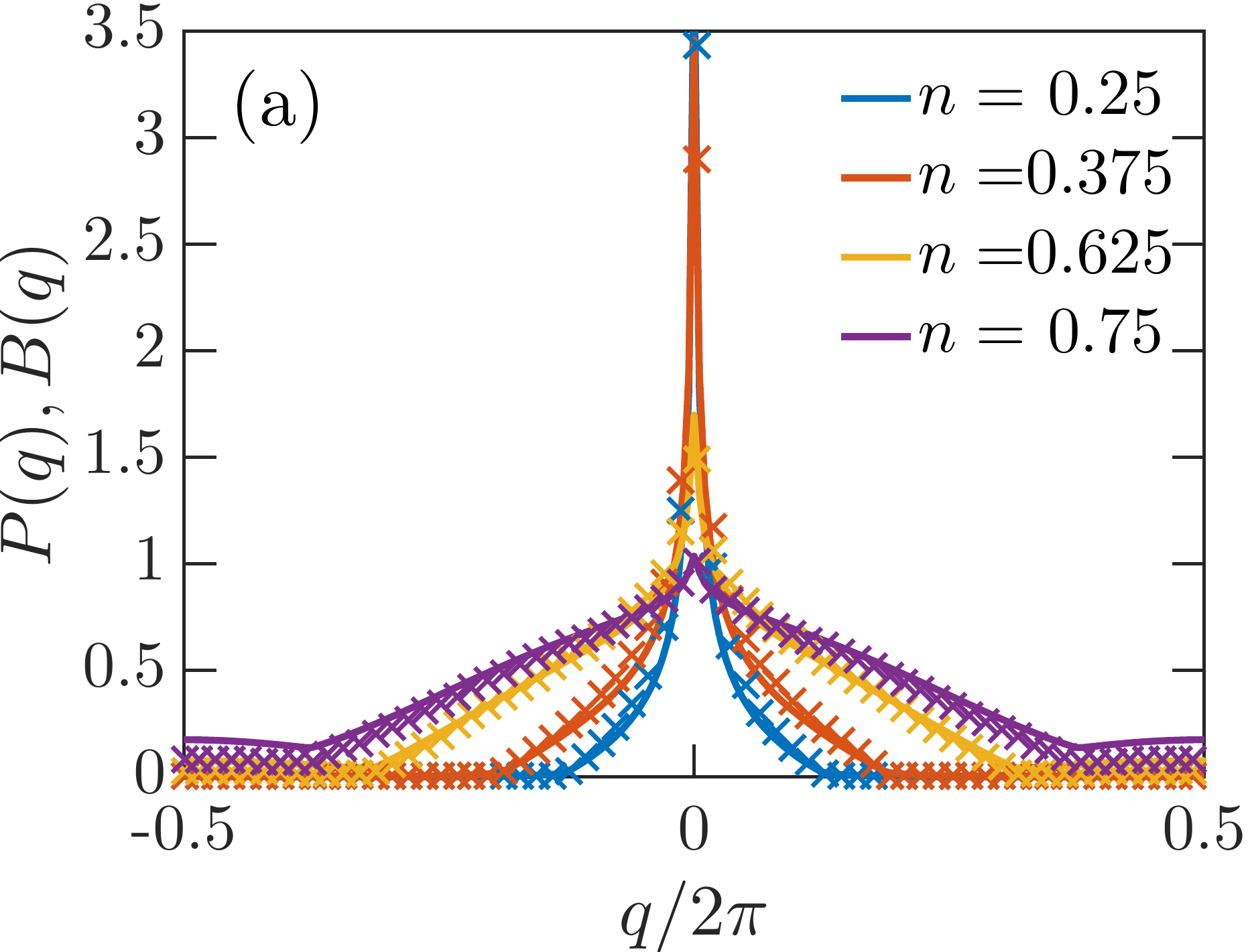}
\includegraphics[width=0.32\linewidth]{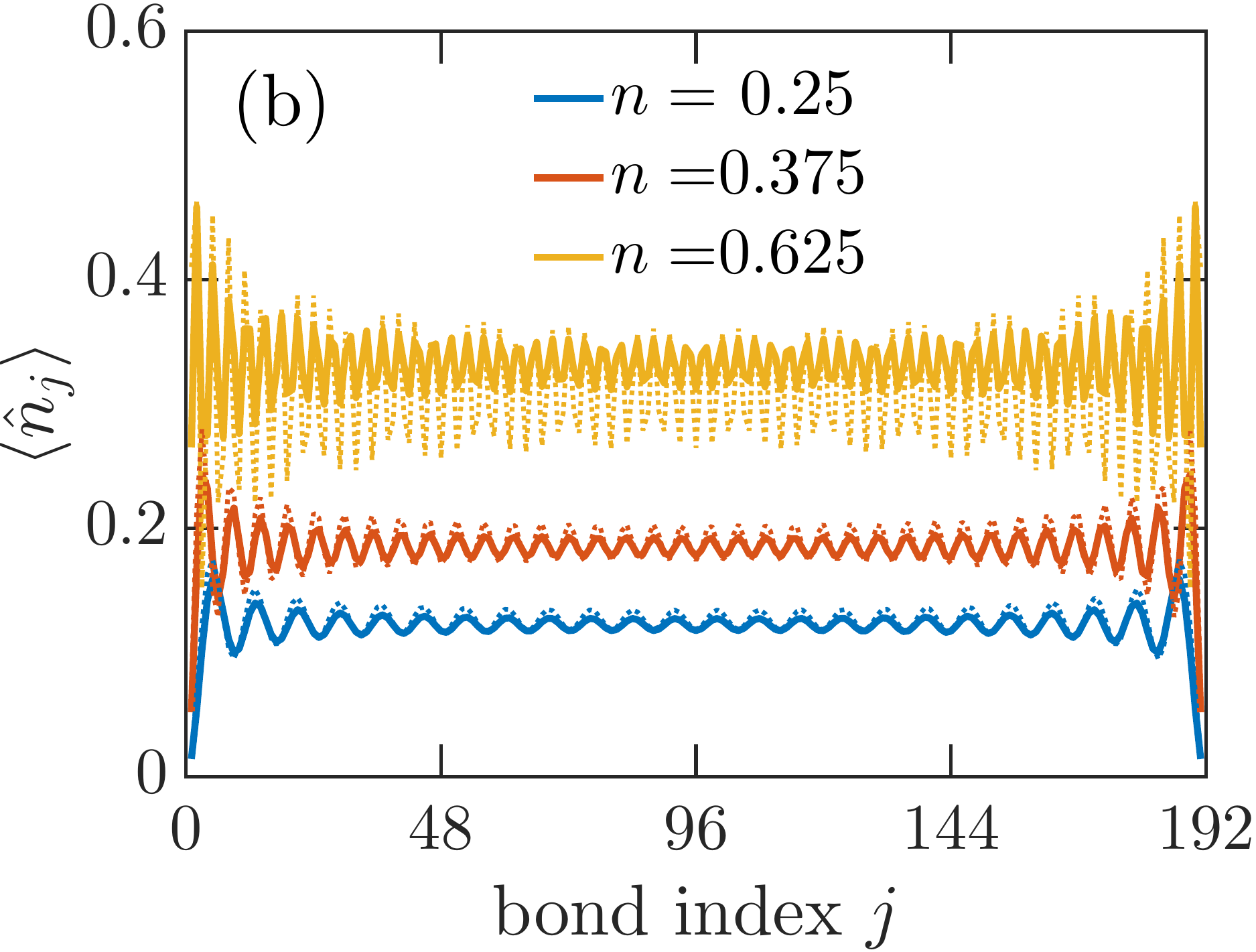}

\includegraphics[width=0.32\linewidth]{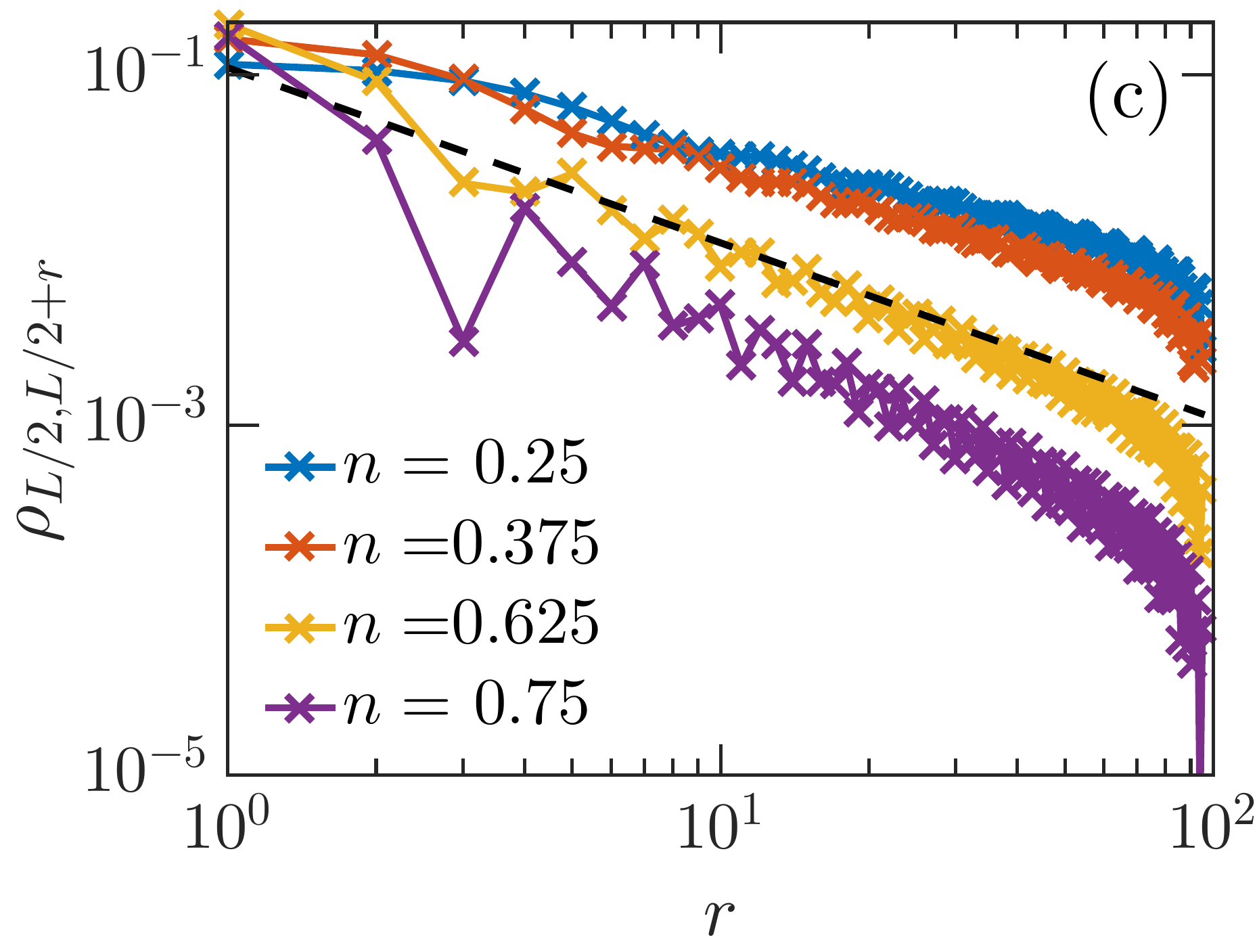}
\includegraphics[width=0.32\linewidth]{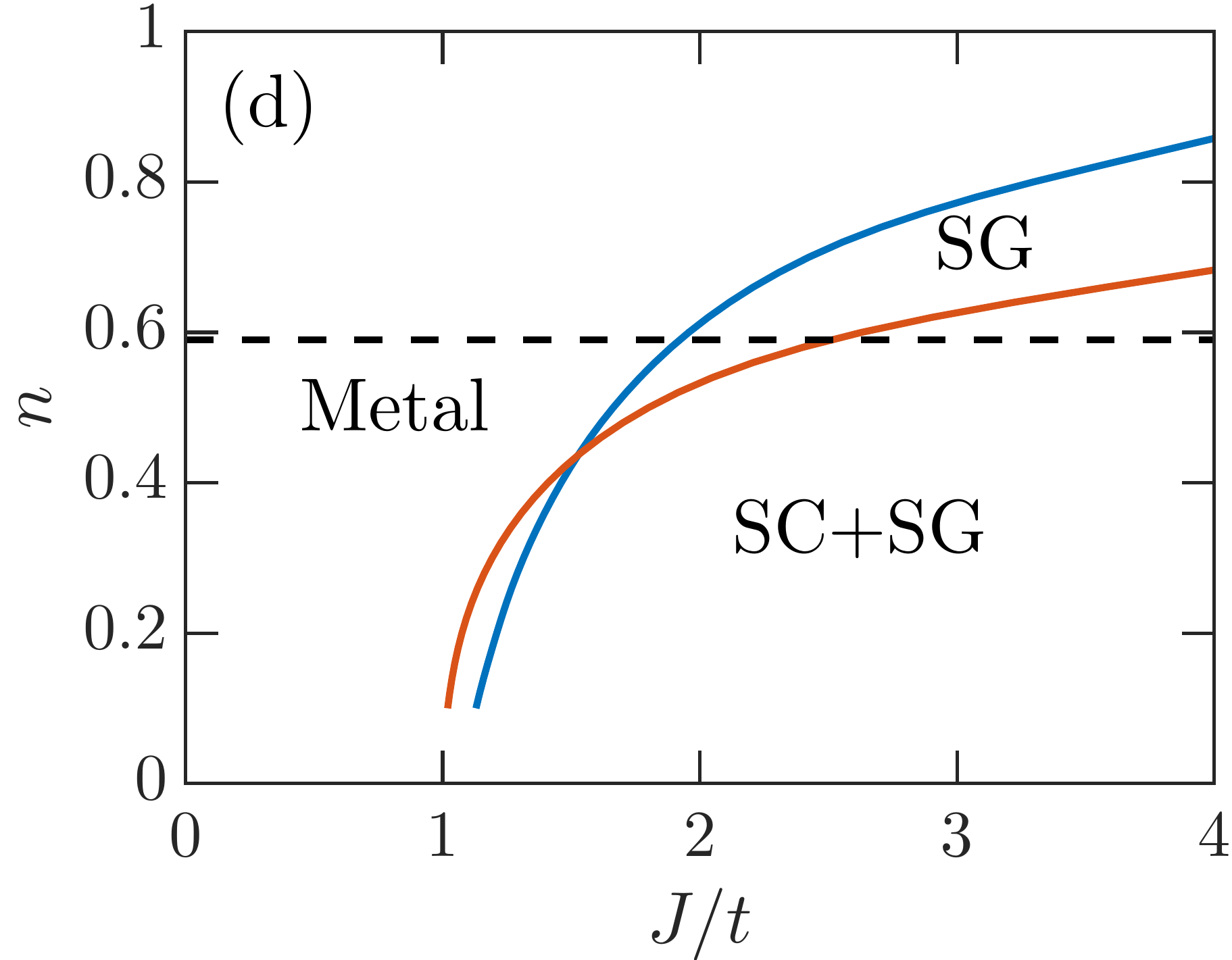}
\end{center}
\caption{A comparison between the $t$--$J$--$\alpha$ model and the projected boson model. $t$--$J$--$\alpha$ model with $J/t = 5$, and $\alpha = 1/2$ computed with $L = 192$ lattice sties. The singlet structure factor $P(q)$ is shown in (a) as the solid line, with the boson structure factor $B(q)$ shown with crosses. We compare the real space densities of singlets and bosons in (b), with the singlets shown with the solid line, and the bosons marked with the dotted line. In (c) a slice of the single boson density matrix as a function of separation $r$ is marked with crosses, with solid lines as a guide to the eye. For reference, a black dashed line indicates the line $0.11r^{-1}$. In (d) we duplicate the $t$--$J$--$\alpha$ model phase diagram at $\alpha = 0.5$ from \fir{fig:phasediagrams}, but now include the superfluid phase boundary of the effective boson model marked as a black dotted line.
\label{fig:BFComparison}}
\end{figure*}

We have so far demonstrated that a finite pair-hopping $\alpha > 0$ leads to suppressed superconductivity at large fillings, coinciding with an increased spin-gap region. To better understand this observation, we now look at the spin-gapped region in more detail. Inside this region, we expect all fermions to be bound into singlet pairs. Given this, we define a new Hilbert space for the system consisting of $L-1$ “sites", which represent the bonds of the original lattice, as illustrated in \fir{fig:BosonModel}(a). These “sites" may be occupied (or not) by a boson representing a singlet pair in the spin-gapped $t$--$J$--$\alpha$ model and thus we have $nL/2$ bosons in the system. The operators $\hat{a}^{\dagger}_{j}$ and $\hat{a}_{j}$ create and annihilate hardcore bosons on “site" $j$, respectively, while $\hat{m}_{j} = \hat{a}^{\dagger}_{j} \hat{a}_{j}$ is the corresponding bosonic number operator.

However, not all configurations of this effective lattice model represent valid configurations in the $t$--$J$--$\alpha$ model, as illustrated in \fir{fig:BosonModel}(b). Since the $t$-$J$--$\alpha$ model does not allow double-occupations, two nearest-neighbour singlet pairs cannot overlap. This manifests itself in the effective bosonic model as the constraint that we cannot have two adjacent “sites" occupied by hardcore bosons. This constraint fortuitously prevents inconsistencies which would arise due to the singlet creation operators $b^{(\dagger)}_{j,j+1}$ not obeying bosonic commutation relations when the singlets overlap. Rather, states which would reveal the composite nature of the bosons are projected out. This effective boson model is closely related to quantum lattice dimer models \cite{Rokhsar1988} thought to have relevance to high-$T_c$ superconductivity in two dimensions. In the limit $J/t \rightarrow \infty$, the Hamiltonian for the effective model is
\begin{equation} \label{eqn:effboson}
\hat{H}_{\rm eff} = \hat{P}_{\rm nn} \left[ -J \sum_{j} \hat{m}_{j} -\alpha J \sum_{j} \left(\hat{a}^{\dagger}_{j} \hat{a}_{j+1} + {\rm H.c.} \right) \right] \hat{P}_{\rm nn}.
\end{equation}
Here, $\hat{P}_{\rm nn}$ is a projection operator which removes states from the Hilbert space which contain bosons on two adjacent sites. This can be seen as a Hamiltonian for hardcore bosons hopping on a lattice with an infinite nearest-neighbour repulsion. Since the total number of bosons is conserved, and we are working at a fixed filling fraction $n$, the first term is a constant $nLJ/2$, and thus can be ignored. The parameter $\alpha J$ then just rescales the energies and does not modify the ground state. This leaves the filling fraction $n$ as the only free parameter in the model. Despite the restriction $\hat{P}_{\rm nn}$ on the hopping, we still expect the bosons to be able to quasi-condense into a superfluid state when the filling is sufficiently small, $n < n_{\rm crit}$. This superfluid of bosons then corresponds to spin-gapped superconductivity in the ${t\textrm{--}J\textrm{--}\alpha}$ model.

The boson structure factor $B(q)$ is \eqr{eq:structurefactor} applied to the single particle density matrix (SPDM) $\rho_{jk} = \langle \hat{a}^{\dagger}_{j} \hat{a}_{k} \rangle$, and is essentially the momentum distribution of bosons. This is shown as crosses in \fir{fig:BFComparison}(a), where we see that when the filling is small, the bosons do not see the extremely strong local repulsive interaction, and so they macroscopically occupy the $q=0$ quasi-momentum state. However, as the filling increases, the repulsive interaction plays a stronger role, and the peak broadens as the bosons are forced, by the interactions, to occupy higher momentum states (quantum depletion of the quasi-condensate). In the same figure, we compare this with the singlet structure factor of the $t$--$J$--$\alpha$ model with $J/t = 5$ (solid lines), finding a very close agreement between the two. 

Similarly, the real-space boson density is shown in \fir{fig:BFComparison}(b). At smaller fillings, small oscillations at a frequency $\pi n$ can clearly be seen. This is because in one-dimension, hardcore bosons inherit the Friedel oscillations from the corresponding Jordan-Wigner fermions \cite{Rigol2005}. Once again, these closely match the oscillations in the real-space singlet density in the $t$--$J$--$\alpha$ model ground state. 


Whether a bosonic lattice system is superfluid or not is determined by the decay of off-diagonal elements in the SPDM. To determine the critical filling of the superfluid transition, we now look at the Luttinger parameter for bosons $K_{b}$ \cite{Ejima2011}. One can show that for $r \gg 1$, the long range behaviour of the SPDM is
\begin{equation}
\rho_{0, r} \sim r^{-K_{b}/2},
\end{equation}
which we show in \fir{fig:BFComparison}(c). As in the fermionic case, we extract $K_{b}$ from the bosonic density structure factor,
\begin{equation}
M(q) \approx \frac{|q|}{2 \pi K_{b}} \quad {\rm as}, \quad q \rightarrow 0,
\end{equation}
where $M(q)$ is \eqr{eq:structurefactor} applied to the correlation function $$M_{j, k} = \langle \hat{m}_{j} \hat{m}_{k} \rangle - \langle \hat{m}_{j} \rangle \langle \hat{m}_{k} \rangle.$$ Computing $K_{b}$ as a function of $n$, we find that the critical filling $n_{\rm crit} \approx 0.59$, which we mark on a copy of the $\alpha = 0.5$ $t$--$J$--$\alpha$ phase diagram, showing a qualitative aggreement for the superconductor-preformed pair transition at large $J/t$. 

In the vicinity of the $t$--$J$--$\alpha$ phase diagram where the effective boson model is valid (i.e. the region with a significant spin-gap), this number provides an estimate of the largest filling at which one can have superconductivity. We expect the single-particle hopping $t$, which we have neglected in the effective boson model, to increase the propensity of the system to superconduct, and so $n_{\rm crit}$ is expected to provide a lower bound on this maximum filling. This is in approximate agreement with the $t$--$J$--$\alpha$ model at $J/t \gg 1$, as we indicate in \fir{fig:BFComparison}(d). Eventually, at maximum filling $n = 1$, the ground state is a (pair) density wave with every other bond being occupied by a hardcore boson. 

\section{Conclusions}\label{sec:conclusions}

We have shown that the effect of the pair-hopping $\alpha J$ in the $t$--$J$--$\alpha$ model is to enhance mobility of pairs, which manifests itself in the ground state phase diagram in a number of ways. Firstly, this pushes the metal-superconducting boundary to lower values of $J/t$ in dilute systems and destabilises the phase-separated region. This has significant implications for periodically-driven Hubbard systems, as it means that driving-induced singlet pairing may be induced at significantly lower strengths than might be expected. Secondly, at larger fillings, superconductivity is suppressed despite the increased pair hopping. We now understand this in the following way: by lowering the energy of bound singlets, the pair-hopping increases the size of the spin-gap region up to much larger values of $n$. Inside the spin-gap region, the physics may be described by a simple model of hardcore bosons with a kinetic constraint whose origins lie in the “no double-occupation" projection of the $t$--$J$ model. Due to these restrictions, the bosons may not condense above a critical filling $n_{\rm crit}$, and so superconductivity in the $t$--$J$--$\alpha$ model cannot occur inside the spin-gap above this filling. This is consistent with the phase diagram, showing a larger spin-gapped region than superconducting region.

It is known that including next-nearest-neighbour hopping terms in one dimensional chains (equivalent to a two-leg ladder system) significantly increases the size of the spin-gap region. This raises the intriguing possibility that kinetic constraints in the spin-gapped phase might play an important role in the physics of high-$T_c$ superconductivity. It is likely that this behaviour may be clearly observed in cold atom experiments, where superexchange physics can be more directly probed \cite{Desbuquois2017, Gorg2018}. In short, studying the $t$--$J$--$\alpha$ model in higher dimensions could provide significant insights into the behaviour of high-$T_c$ superconductors and periodically driven strongly correlated systems. 

\section*{Acknowledgements}
This research is funded by the European Research Council under the European Union's Seventh Framework Programme (FP7/2007--2013)/ERC Grant Agreement no.~319286 Q-MAC. D.J. acknowledges support from the EPSRC under grant Nos. EP/K038311/1 and EP/P009565/1. S.R.C. gratefully acknowledges support from the EPSRC under grant No. EP/P025110/1.

\appendix

\section{Floquet engineering the $t$--$J$--$\alpha$ model} \label{sec:app_floquet}
\begin{figure*}
\begin{center}
\includegraphics[width=0.32\linewidth]{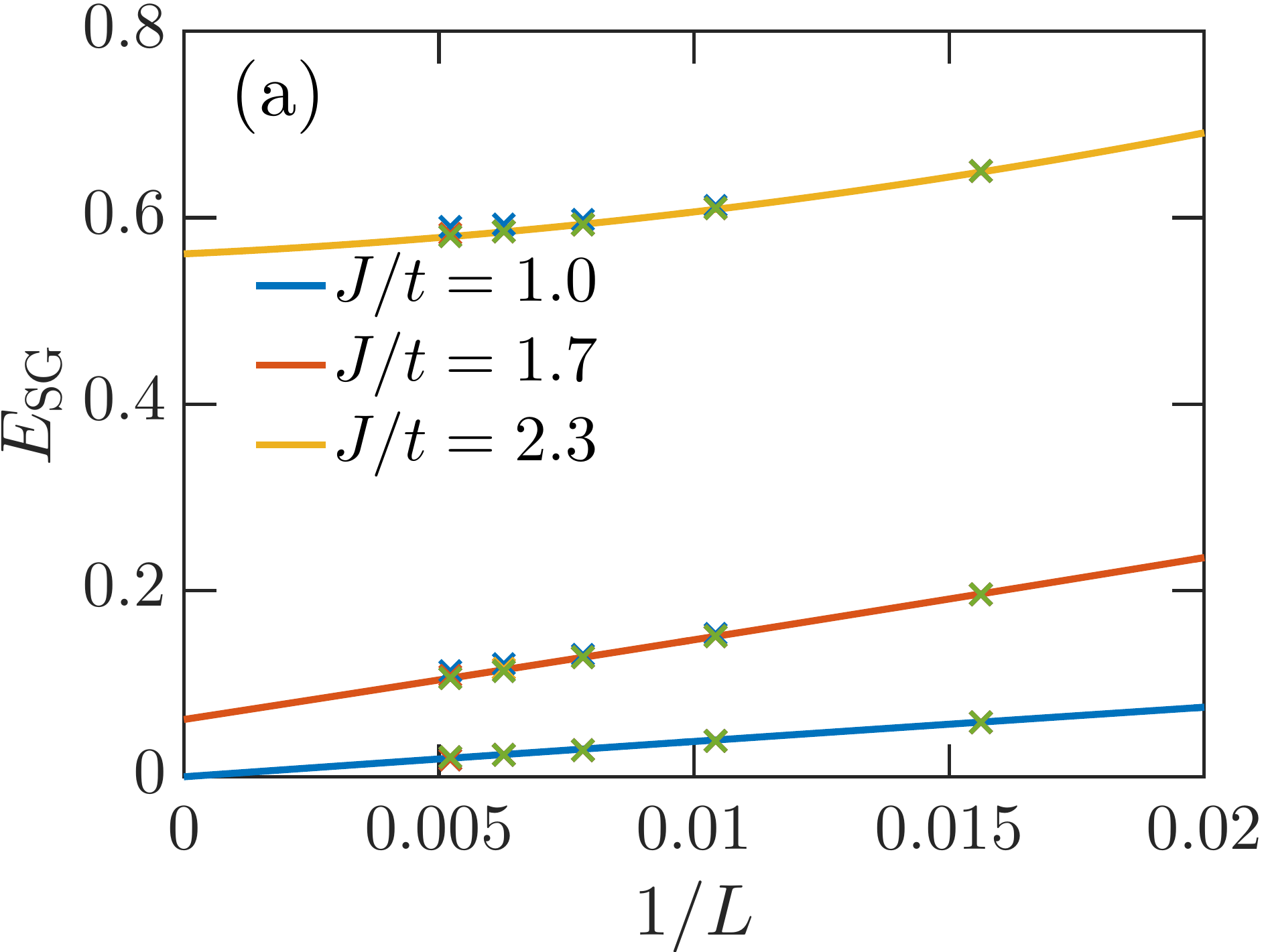}
\includegraphics[width=0.32\linewidth]{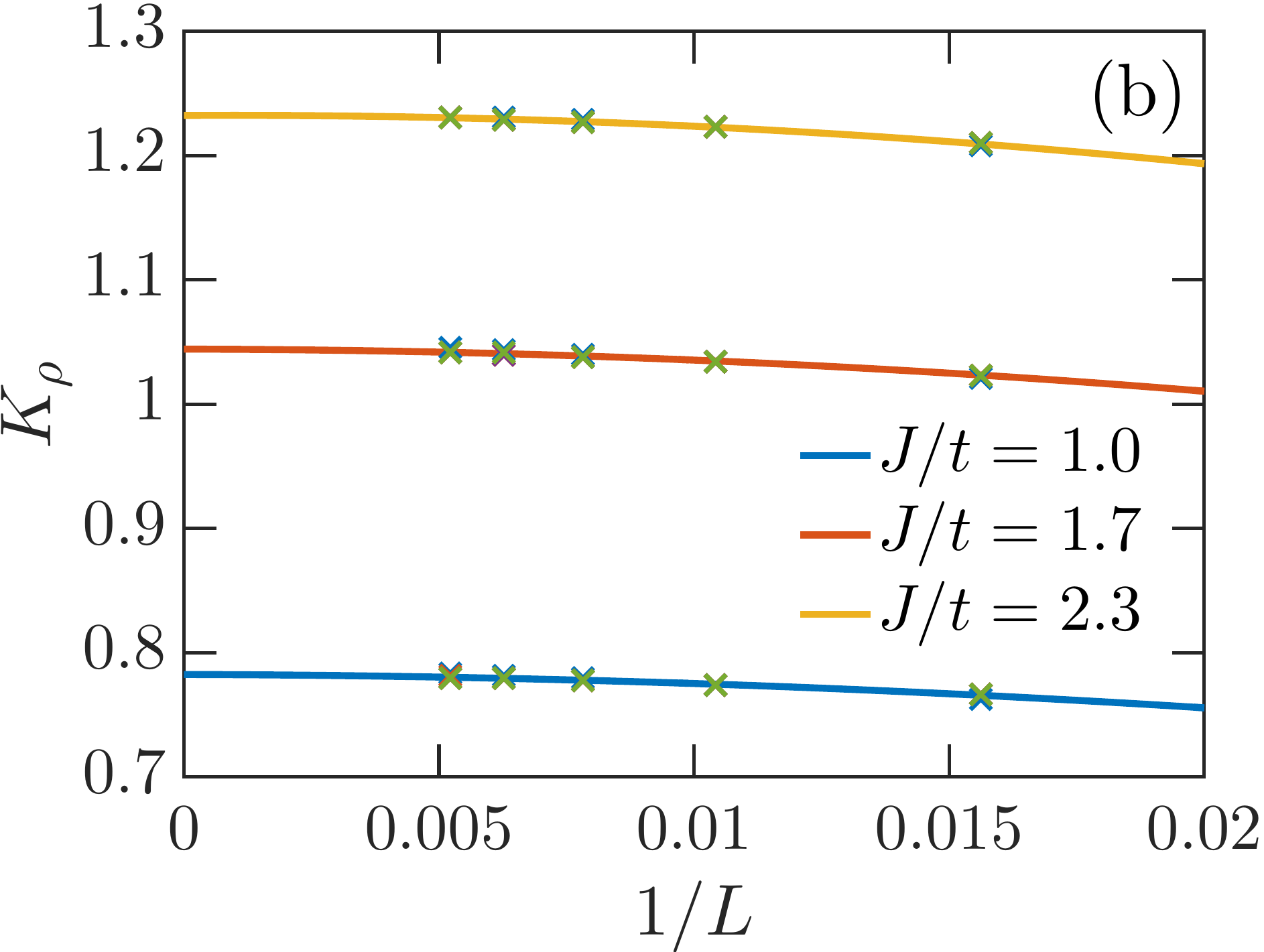}
\includegraphics[width=0.31\linewidth]{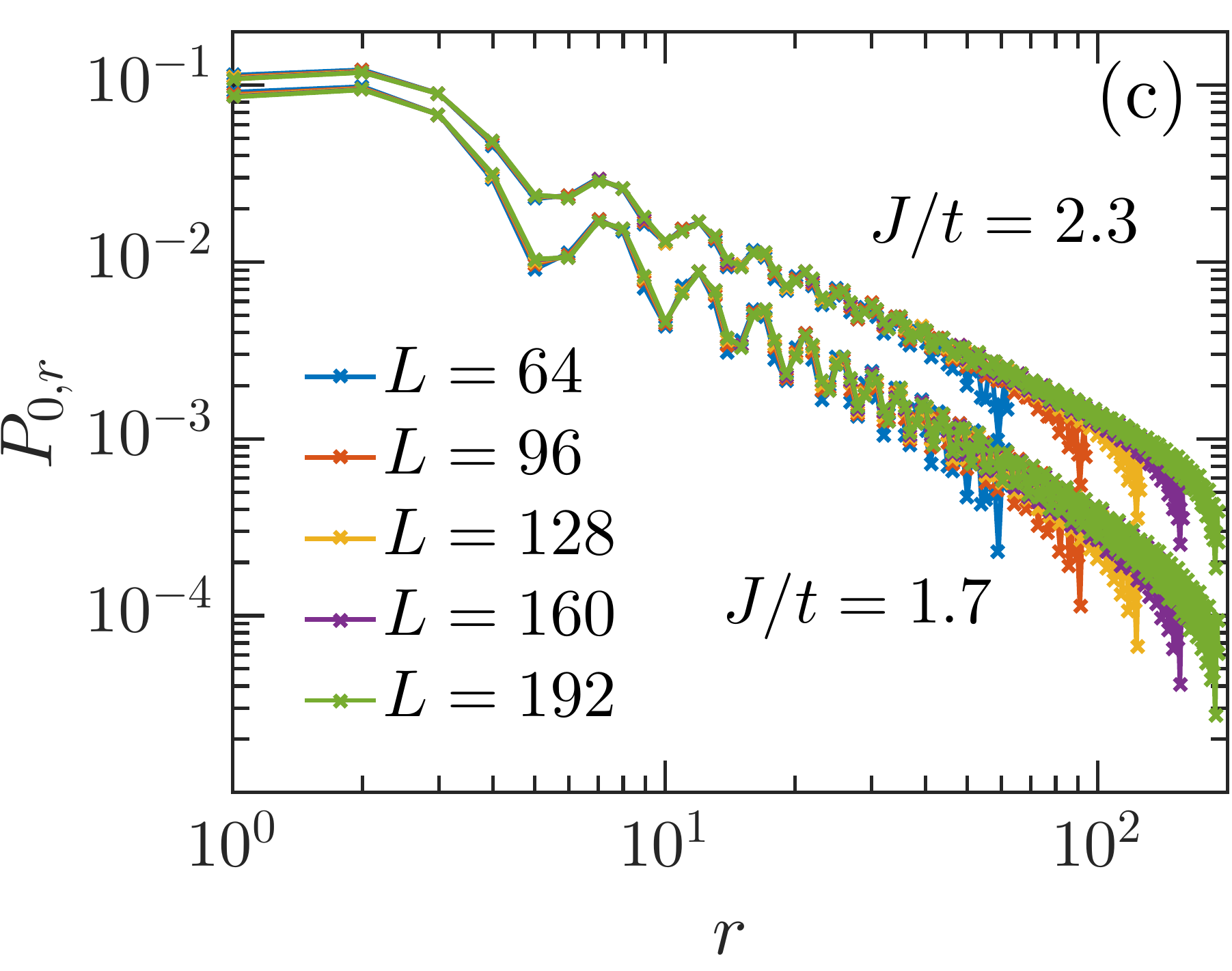}
\end{center}
\caption{Examples of the finite size extrapolation, the system parameters $\alpha = 0.5$, $n = 7/32$, with $J/t$ marked on the plots. In (a) and (b) we show the extrapolation of the spin gap and Luttinger Parameter respectively, as a function of system size. The crosses (which overlap) are computed for bond dimensions $\chi = 100-500$. In (c) we show the singlet correlations at selected interaction strengths with increasing system size computed for $\chi = 300$. 
\label{fig:Extrapolation}}
\end{figure*}

Here we outline how one can engineer the $t$--$J$--$\alpha$ model out of equilibrium by periodically driving a Hubbard model. Possible experimental implementations include, for instance, shaking an optical lattice \cite{Desbuquois2017, Gorg2018}, or driving vibrations in an organic solid \cite{Kaiser2014, Singla2015, Coulthard2017}. We begin with a one-dimensional single-band Hubbard model,
\begin{equation}
\hat{H}_{\rm Hub} = U \sum_{j} \hat{n}_{j, \uparrow} \hat{n}_{j, \downarrow} - t \sum_{j, \sigma} \left(\hat{c}^{\dagger}_{j, \sigma} \hat{c}_{j+1, \sigma} + {\rm H.c.} \right).
\end{equation}
We then add a periodic driving term
\begin{equation}
\hat{H}_{\rm drive}(\tau) = V \sin(\Omega \tau)  \sum_{j} j \hat{n}_{j}.
\end{equation}
This particular driving term models a cloud of ultracold atoms trapped in an optical lattice, where the lattice itself is shaken with an angular frequency $\Omega$, or an AC “electric field" is applied across the system. However, similar physics is shown to occur with other driving terms, such as those induced by a travelling wave \cite{Bukov2015}, or where odd and even sites experience different driving strengths \cite{Coulthard2017}.

As the Hamiltonian $\hat{H}_{\rm Hub} + \hat{H}_{\rm drive}(\tau)$ is periodic in time, we are able to use Floquet theory \cite{Shirley1965, Casas2001, Eckardt2015} to compute an effective static Hamiltonian for the stroboscopic evolution of the system. In this instance we focus on the far off-resonant, in-gap regime $t \ll \Omega \ll U$. We seek an effective Hamiltonian which describes the low-energy physics of the Hamiltonian, which we obtain via a generalised Schrieffer--Wolff transformation (SWT) \cite{Bukov2016}. The dynamics generated by the Hamiltonian will contain oscillations at frequencies $\Omega$ and $U$, both of which are large compared to $t$. The SWT amounts to a sequence of rotating wave approximations where we systematically eliminate frequencies from highest to lowest. 

We begin by performing the standard SWT to order $t/U$ to obtain the $t$--$J$--$\alpha$ model
\begin{eqnarray}
\hat{H}_{tJ \Omega} &=& \hat{P} \bigg[ -t_0 \sum_{j, \sigma} \left( c^{\dagger}_{j, \sigma} c_{j+1, \sigma} + {\rm H.c.} \right) \nonumber \\
& & - \frac{t_0^2}{U} \sum_{j} \hat{b}_{j,j+1}^{\dagger} \hat{b}_{j,j+1} + \frac{t_0^2}{2U} \left( \hat{b}^{\dagger}_{j, j+1} \hat{b}_{j+1, j+2} + {\rm H.c.} \right) \nonumber \\ 
& & + V \cos(\Omega \tau)  \sum_{j} j \hat{n}_{j} \bigg] \hat{P}, 
\end{eqnarray}
where $\hat{P}$ is a projector onto state which contain no double-occupations. From here we transform into the rotating frame with respect to the driving term, and perform a high-frequency Magnus expansion \cite{Bukov2015} to obtain the effective Hamiltonian
\begin{eqnarray}
\hat{H}_{tJ \alpha} &=& \hat{P} \bigg[ - \mathcal{J}_{0}\left(\frac{V}{\Omega}\right) t_0 \sum_{j, \sigma} \left( c^{\dagger}_{j, \sigma} c_{j+1, \sigma} + {\rm H.c.} \right) \nonumber \\
& & - \frac{t_0^2}{U} \sum_{j} \hat{b}_{j,j+1}^{\dagger} \hat{b}_{j,j+1} \nonumber \\
& & - \mathcal{J}_{0}\left(\frac{2V}{\Omega}\right) \frac{t_0^2}{2U} \left( \hat{b}^{\dagger}_{j, j+1} \hat{b}_{j+1, j+2} + {\rm H.c.} \right) \bigg] \hat{P}. \nonumber \\ 
\end{eqnarray}
We now identify this as the Hamiltonian in \eqr{eqn:tjsinglet}, with $t = \mathcal{J}_{0}(V/\Omega) t_0$, $J = t_0^2/U$, and $\alpha = \mathcal{J}_{0}(2V/\Omega)$. In an optical lattice context, where one has a fine degree of control over all parameters $t_0$, $U$, $\Omega$, and $V$, one can semi-independently vary $t$, $J$, and $\alpha$, and explore experimentally the effect of pair-hopping on superconductivity in higher dimensions. 

We note that this procedure is valid only when there is no “beating"  between the oscillations at frequencies $\Omega$ and $U$. In other words, we require $|U-\Omega| \gg t$. The method can be generalised to the near-resonant case by simultaneously eliminating the driving term along with an amount $\Omega$ of the interaction term, leaving a Hamiltonian with an effective on-site repulsion $U-\Omega$, as discussed in \cite{Bukov2016, Desbuquois2017}.




\section{Details of the DMRG calculation} \label{sec:dmrg_details}

In this section we summarise some technical details of the DMRG calculation. The advantage of using a finite-sized algorithm rather than infinite-DMRG is that we may use symmetries to exactly fix the number of fermions in the system $N_{\uparrow}$ and $N_{\downarrow}$, which allows the precise determination of quantities such as the spin gap and compressibility. 

The drawback of studying such a finite system with open boundaries is that it requires us to consider the interplay between finite-size and finite-entanglement scaling \cite{Dolfi2015}. As we detail in the following, we find that our results are dominated by finite-size effects rather than finite-entanglement artefacts, and so we extrapolate only to $L \rightarrow \infty$ for the largest value of $\chi$ used. 

We show some typical finite-size extrapolations in \fir{fig:Extrapolation}(a) and (b). In (a), we show the spin gap at $\alpha = 0.5$, $n = 7/32$ as a function of inverse system size $1/L$ for various values of $J/t$. The solid lines are quadratic fits for different interaction strengths. The crosses are data points for different system sizes and values of $\chi = 100-500$  (increasing $\chi$ makes almost no difference to the results and thus the multiple crosses appear as a single cross). Similarly, we show the extrapolation of the Luttinger parameter in \fir{fig:Extrapolation}(b), where the lines and crosses have the same meaning as in (a).

In \fir{fig:Extrapolation}(c), we show the nearest-neighbour singlet correlation function at various system sizes for the same parameters as in plots (a) and (b), with $\chi = 300$. We see a polynomial decay over a significant range of distances $r$, after which an exponential tail develops. Such exponential tails always appear due to finite size and finite entanglement in some combination. The value of $r$ at which this exponential tail sets in increases as we increase the system size. at this bond dimension, finite-size effects dominate over finite-entanglement effects, and thus we extrapolate only in $L$ and not $\chi$ \cite{Dolfi2015}.

In all phase diagrams in this paper, we compute the ground state at intervals of $\Delta J = 0.1$, and $\Delta n = 1/16$. For each quantity which determines a phase boundary, a linear interpolation is performed at the boundary of these grid squares to obtain a set of approximate grid points for the phase boundary. We then interpolate these points with a smoothed cubic spline.

\bibliography{PhaseDiagram}

\end{document}